\documentclass[%
 reprint,
 amsmath,amssymb,
 aip,
]{revtex4-2}

\usepackage[T1]{fontenc}
\usepackage{upgreek}

\usepackage{multirow}
\usepackage{makecell}
\usepackage{booktabs}
\usepackage{colortbl}
\usepackage{graphicx}% Include figure files
\usepackage{dcolumn}% Align table columns on decimal point
\usepackage{bm}% bold math
\usepackage{subcaption}
\usepackage{siunitx}
\usepackage{pifont}% http://ctan.org/pkg/pifont

\DeclareMathOperator*{\argmin}{arg\,min}

\newcommand{\taur}{\tilde{\tau}}

\makeatletter
\def\@email#1#2{%
 \endgroup
 \patchcmd{\titleblock@produce}
  {\frontmatter@RRAPformat}
  {\frontmatter@RRAPformat{\produce@RRAP{*#1\href{mailto:#2}{#2}}}\frontmatter@RRAPformat}
  {}{}
}%
\makeatother

\begin{document}

\title{Transfer Learning and the Early Estimation of Single-Photon Source Quality using Machine Learning Methods}

\author{David Jacob Kedziora}
%\email{david.kedziora@uts.edu.au}
\affiliation{Complex Adaptive Systems Lab, University of Technology Sydney, Sydney, Australia}

\author{Anna Musiał}
% \email{anna.musial@pwr.edu.pl}
\affiliation{Department of Experimental Physics, Wroclaw University of Science and Technology, Wrocław, Poland}

\author{Wojciech Rudno-Rudziński}
% \email{wojciech.rudno-rudzinski@pwr.edu.pl}
\affiliation{Department of Experimental Physics, Wroclaw University of Science and Technology, Wrocław, Poland}

\author{Bogdan Gabrys}
\email{bogdan.gabrys@uts.edu.au}
\affiliation{Complex Adaptive Systems Lab, University of Technology Sydney, Sydney, Australia}

\date{\today}

\begin{abstract}
The use of single-photon sources (SPSs) is central to numerous systems and devices proposed amidst a modern surge in quantum technology.
However, manufacturing schemes remain imperfect, and single-photon emission purity must often be experimentally verified via interferometry.
Such a process is typically slow and costly, which has motivated growing research into whether SPS quality can be more rapidly inferred from incomplete emission statistics.
Hence, this study is a sequel to previous work that demonstrated significant uncertainty in the standard method of quality estimation, i.e.~the least-squares fitting of a physically motivated function, and asks: can machine learning (ML) do better?
The study leverages eight datasets obtained from measurements involving an exemplary quantum emitter, i.e.~a single InGaAs/GaAs epitaxial quantum dot; these eight contexts predominantly vary in the intensity of the exciting laser.
Specifically, via a form of `transfer learning', five ML models, three linear and two ensemble-based, are trained on data from seven of the contexts and tested on the eighth.
Validation metrics quickly reveal that even a linear regressor can outperform standard fitting when it is tested on the same contexts it was trained on, but the success of transfer learning is less assured, even though statistical analysis, made possible by data augmentation, suggests its superiority as an early estimator.
Accordingly, the study concludes by discussing future strategies for grappling with the problem of SPS context dissimilarity, e.g.~feature engineering and model adaptation.
\end{abstract}

% \keywords{quantum dots; quantum communication; emission statistics; bootstrapping; generative models; uncertainty analysis}

\maketitle

\section{Introduction}
\label{Sec:Intro}

The efficient design and fabrication of a reliable on-demand single-photon source (SPS) is a core research agenda within the field of modern quantum optics.
Indeed, the ability to produce individual photons with precise control, leveraging their quantum nature, is a crucial enabler for various other technologies in quantum communication~\cite{kimb08, ko10}, cryptography~\cite{safu11, benn14} and computation~\cite{kira04, zhwa20, zhma22, yama05, cout23}.
Thus, unsurprisingly, this research endeavour has attracted plenty of effort and attention since the turn of the century, when the first SPS devices based on quantum dots were realised~\cite{kuma00, miki00}.

However, engineering an SPS does not come without challenges.
Fabrication methods are never ideal, whether involving colloidal growth~\cite{chte17, zhma22}, epitaxy~\cite{yuka02, mi17, scre21}, or crystalline defects~\cite{doma13, hecl15, hebe15, safo20, bavi21, gahe23}.
Even once the SPS is fabricated, its operation is dependent on the characteristics of the laser exciting the quantum dot, especially on the excitation scheme applied~\cite{Huber2015, Doris}.
It is also sensitive to many environmental factors, such as temperature and electromagnetic noise.
Quality control is thus a vital part of SPS engineering.

The fundamental marker for SPS quality is single-photon emission purity, and the corresponding `impurity' is experimentally determined by the probability of multi-photon emission (MPE), which is the focus of this research article.
Specific applications may care about many other factors, such as polarisation control or photon indistinguishability, but these are typically secondary to ensuring the proportion of MPE events is below a requisite threshold.
In practice, such purity is typically measured by beam-splitting SPS emission and applying `Hanbury Brown and Twiss' (HBT) interferometry~\cite{brtw56} to experimentally determine a second-order auto-correlation function on photon intensity~\cite{kida77, kuma00, zwbl01, buri12}, commonly referred to as $g^{(2)}(\tau)$.
Specifically, calibrated observations of a photon hitting each detector at the same time indicate that MPE has occurred, i.e.~$g^{(2)}(0)\neq 0$.
Importantly, a threshold of $g^{(2)}(0)=0.5$ defines an SPS, marking where single-photon emission is more common than MPE, but state-of-the-art (SOTA) engineering efforts often aim for $g^{(2)}(0)$ values that are orders of magnitude smaller~\cite{mita16, hafi18, scjo18}.

The measurement of temporal photon correlations has applications beyond just the determination of MPE probability, as it is used, for instance, to determine the level of photon entanglement~\cite{lica17, scre21, de20}; it is thus central to experimental quantum optics. Normalised correlation functions give insight into the photon-number properties of light. The higher-order moments allow both fast state classification and its in-depth characterisation~\cite{cor1}, benefitting from the fact that the results are typically insensitive to experimental losses. Recent progress in detector technologies has further extended the use of photon correlation techniques to imaging methods, where the resolution can be improved by taking advantage of non-classical light properties, including entangled-photon illumination schemes~\cite{cor2}.

The common problem for all these techniques is that measuring correlation is usually a resource-expensive process. Taking SPS as an example, modern SPS technologies exhibiting the lowest probability of MPE, especially SOTA devices, need to be run at costly cryogenic temperatures.
Moreover, the separate detectors in an HBT interferometer typically require a low photon flux to resolve individual photons; it can take hours to build up an accurate picture of $g^{(2)}(\tau)$.
Thus, in recent years, there has been deepening contemplation on whether SPS quality can be estimated quickly from limited data.
Such efforts have primarily examined ways of fitting theoretical forms of $g^{(2)}(\tau)$ to small-sample histograms~\cite{coad20}.

Unfortunately, using the technique of data augmentation, new research has highlighted just how substantial and unavoidable the uncertainty in early estimates based on fitting can be~\cite{kemu23}.
Even so, if experimentalists are happy to embrace a certain level of uncertainty, a question remains: is fitting a small sample of two-photon detections the best way to estimate $g^{(2)}(0)$?
In grappling with this question, some might consider the predictive techniques of machine learning (ML), which have increasingly cross-pollinated various disciplines in the physical sciences to great effect.
However, it is unclear how to cast the early estimation of SPS quality as an ML problem.
After all, upon encountering data from a new SPS `experimental context' for the first time, there are no `ground-truth' values of target variable $g^{(2)}(0)$ to supervise the training of an ML model.
By the time $g^{(2)}(0)$ is determined for an SPS within an experimental context, an ML model trained on this value is no longer necessary.

That said, what if the lessons machine-learned for one experimental context could then be \textit{transferred} to another?
This paper explores this idea, examining ML techniques applied to data from a single epitaxial InGaAs/GaAs quantum dot within a deterministic photonic nanostructure that emits at 1.3 \si{\um}~\cite{muzo20}.
The only physical factor that changes across eight datasets related to this SPS is the intensity of the exciting laser, which influences $g^{(2)}(0)$, allowing for a controllable investigation into the merits of transfer learning.
It is ultimately found that, in terms of ML models applied to the \textit{same} contexts they were trained on, even a linear regressor can outperform least-squares fitting.
However, the benefits of \textit{transfer} learning are much more ambiguous, despite data augmentation enabling a statistical analysis that suggests its overall superiority.
Essentially, context dissimilarity remains a challenge, even in this case of seemingly minute variations, and supplementary strategies will likely be needed to support model transferability, e.g.~adaptation.

The paper is organised as follows.
Section~\ref{Sec:Theory} details the expected emission statistics for a quantum dot excited by a pulsed laser, i.e.~what a histogram of two-photon coincidences detected by HBT interferometry should look like in the long term.
Section~\ref{Sec:Methodology} then describes the eight datasets analysed in this work (\ref{Sec:Datasets}) and the two forms of early estimation for SPS quality that are being compared (\ref{Sec:Estimators}).
Specifically, one form derives $g^{(2)}(0)$ from the least-squares fitting of a theoretical formula to a coincidence histogram (\ref{Sec:Fitting}), while the other form involves the `hyperparameter' optimisation and training of five linear/ensemble ML models; this is done on seven SPS contexts prior to deployment on an `unseen' eighth (\ref{Sec:ML}).
Section~\ref{Sec:Results} presents the results of the comparison, first examining ML models trained/tested only on experimental data (\ref{Sec:Experiment}), then statistically analysing ML models trained/tested on synthetic data (\ref{Sec:Synthetic}), and concluding by exploring whether adaptation can further improve performance (\ref{Sec:Adaptive}).
Section~\ref{Sec:Discussion} subsequently acknowledges the challenge of context dissimilarity and discusses future strategies for improving ML model transferability.
Finally, Section~\ref{Sec:Conclusion} summarises the conclusions of this research.

\section{Theory}  
\label{Sec:Theory}

The probability of MPE is generally measured by HBT interferometry.
In such a setup, two single-photon detectors examine beam-split emission and identify moments where both trigger near-simultaneously, referred to here as `two-photon events' or `coincidences'.
Each detected event is associated with a time delay between the two triggers, $\tau$, although, due to length differences in both optical paths and electrical connections for a real interferometer, this needs to be calibrated from a raw value by an offset, i.e.~$\tau = \taur - \tau_0$.
Any events occurring with a raw time delay of $\taur = \tau_0$ are indicative of MPE.
In practice, the analysis of interferometry-based measurements also groups coincidences; index $i$ is introduced here to represent an associated bin with width $\Delta\tau$ centred at a raw time delay of $\taur_i$, related to a calibrated delay of $\tau_i$ in the usual way by offset $\tau_0$. Minimum bin-width $\Delta\tau$ is typically determined by counting electronics; coincidences are indistinguishable within this resolution.

Now, the theoretical shape of emission profiles across the $\taur$ domain is well understood in the field for common contexts, as long as some physical assumptions hold, e.g.~the observation of two-photon events adhering to Poisson statistics.
In the case of an SPS being excited by a pulsed laser with period $\Lambda$, the long-term histogram of accumulated detections should resemble a $\Lambda$-periodic `comb' of peaks, albeit with the MPE peak at $\taur = \tau_0$ being relatively diminished.
Each of these peaks can be modelled as a two-sided exponential function with decay factor $\gamma_p$, while the envelope of the comb has its own decay factor of $\gamma_e$.

In short, the average number of coincidences within bin $i$ after running the interferometry measurement for time $t$ is $N_i = R_i t$, where the average observation rate for bin $i$, i.e.~$R_i = R(\tau_i)$, is defined as
\begin{equation}
\label{Eq:Fit}
    R(\tau_i; \theta) = R_b + R_p e^{-\gamma_e |\tau_i|} \left( g e^{-\gamma_p |\tau_i|} + \sum_{k\neq 0} e^{-\gamma_p |\tau_i-k\Lambda|}\right).
\end{equation}
In this equation, reformulated into a more computationally amenable form elsewhere~\cite{kemu23}, $R_p$ is the peak observation rate, capturing coincidences due to sequential pulse excitation, while $R_b$ is the background rate, predominantly caused by detector dark counts.
Both constants are dependent on $\Delta\tau$, which should be sufficiently small to properly resolve the comb.
As for $g$, this is the indicator of MPE probability, i.e.~$g^{(2)}(0)$, and thus the focus of this research; the short-form variable will be used within the rest of the text for convenience.
Altogether, the seven fixed parameters in this function can be grouped in a set: $\theta = \left\{\tau_0, R_b, R_p, g, \gamma_e, \gamma_p, \Lambda\right\}$.

\section{Methodology}
\label{Sec:Methodology}

The research described in this article revolves around the computational analysis of experimental interferometry measurements. To support reproducibility, the data and Python scripts are available at \href{https://github.com/UTS-CASLab/sps-quality}{github.com/UTS-CASLab/sps-quality}, commit 7c2880c.
Here, the experimental datasets are detailed in Sec.~\ref{Sec:Datasets}, and the methods for estimating $g$ are described in Sec.~\ref{Sec:Estimators}, i.e.~the least-squares fitting of a physically motivated function (Sec.~\ref{Sec:Fitting}) and the use of ML (Sec.~\ref{Sec:ML}).

\subsection{The Datasets}
\label{Sec:Datasets}

As an extension of previous research~\cite{kemu23}, now examining the utility of predictive ML, this work examines the same eight experimental datasets sourced from interferometry-based measurements of a `fibre-coupled semiconductor single-photon source for secure quantum-communication in the 1.3 \si{\um} range' (FI-SEQUR).
They all involve the same transition for the same single InGaAs/GaAs epitaxial quantum dot, which is positioned deterministically with respect to the centre of a photonic mesa structure.
Further fabrication~\cite{srmu18, zomu19} and interferometry~\cite{muzo20} details are available elsewhere.
As for what drives the SPS, the excitation is above-band, involving an 80 MHz pulsed semiconductor laser operating at a wavelength of 805 nm and a pulse length of 50 ps.

\begin{table}[htb!]
\centering
\caption{Descriptive summary of the eight FI-SEQUR datasets.}
\label{Tab:Datasets}
\begin{tabular}{|c|ccccc|}
\hline
\begin{tabular}[c]{@{}c@{}}Shorthand\\ Label\end{tabular} &
  \begin{tabular}[c]{@{}c@{}}Laser\\ Intensity\\ (\si{\uW})\end{tabular} &
  \begin{tabular}[c]{@{}c@{}}Single\\ Detector\\ Counts\\ Per Second\\ (Hz)\end{tabular} &
  \begin{tabular}[c]{@{}c@{}}Total\\ Events\end{tabular} &
  \begin{tabular}[c]{@{}c@{}}Total\\ Duration\\ (s)\end{tabular} &
  \begin{tabular}[c]{@{}c@{}}Average\\ Event\\ Rate\\ (Hz)\end{tabular} \\ \hline
1p2uW & 1.2 & 3000  & 65037 & 23950 & 2.716  \\
2p5uW & 2.5 & 4000  & 51534 & 10510 & 4.903  \\
4uW   & 4   & 4100  & 61246 & 9390  & 6.522  \\
8uW   & 8   & 5100  & 21611 & 2920  & 7.401  \\
10uW-- & 10  & 6000  & 28817 & 3010  & 9.574  \\
10uW+ & 10  & 12000 & 45198 & 1210  & 37.354 \\
20uW  & 20  & 7000  & 55469 & 4780  & 11.604 \\
30uW  & 30  & 7000  & 52088 & 4780  & 10.897 \\ \hline
\end{tabular}
\end{table}

Crucially, Table~\ref{Tab:Datasets} shows that the only difference between the experimental contexts represented by the eight datasets is the intensity of the laser that drives SPS emission.
Granted, the duration that HBT interferometry was run differs between the contexts, resulting in varying quantities of data to analyse.
The experimental setup was also adjusted for the 10uW+ experiment to increase the rate of observations.
Nonetheless, this group of datasets is otherwise ideal for a proof-of-principle investigation of ML, the similar contexts averting confounding factors.

In greater detail, each dataset is a 2D matrix of coincidence counts, binned in 10-second detection `snapshots' along one axis and intervals of time separation along the other.
Specifically, the $\taur$ domain covers raw delays ranging from $0$ ns to approximately $500$ ns, i.e.~$1954$ bins with a $\Delta\tau$ of $0.256$ ns.
Technically, the 1p2uW dataset has $3907$ bins with a $\Delta\tau$ of $0.128$ ns, but pairs of adjacent bins are summed together -- a negligible error arises from the nonexistent $3908$th bin -- so that all datasets have the same `feature space' of $1954$ columns.
Notably, the constrained $\taur$ domain means that, although a single detector may acknowledge thousands of photon counts per second, only a handful of those constitute one half of a coincidence event; see Table~\ref{Tab:Datasets}.
Nevertheless, this chosen range still encompasses the calibrated `zero delay', i.e.~the MPE peak, which occurs at about $60$ ns due to the previously discussed electronic offset.

\begin{figure*}[p!]
\centering
\includegraphics[width=0.95\textwidth]{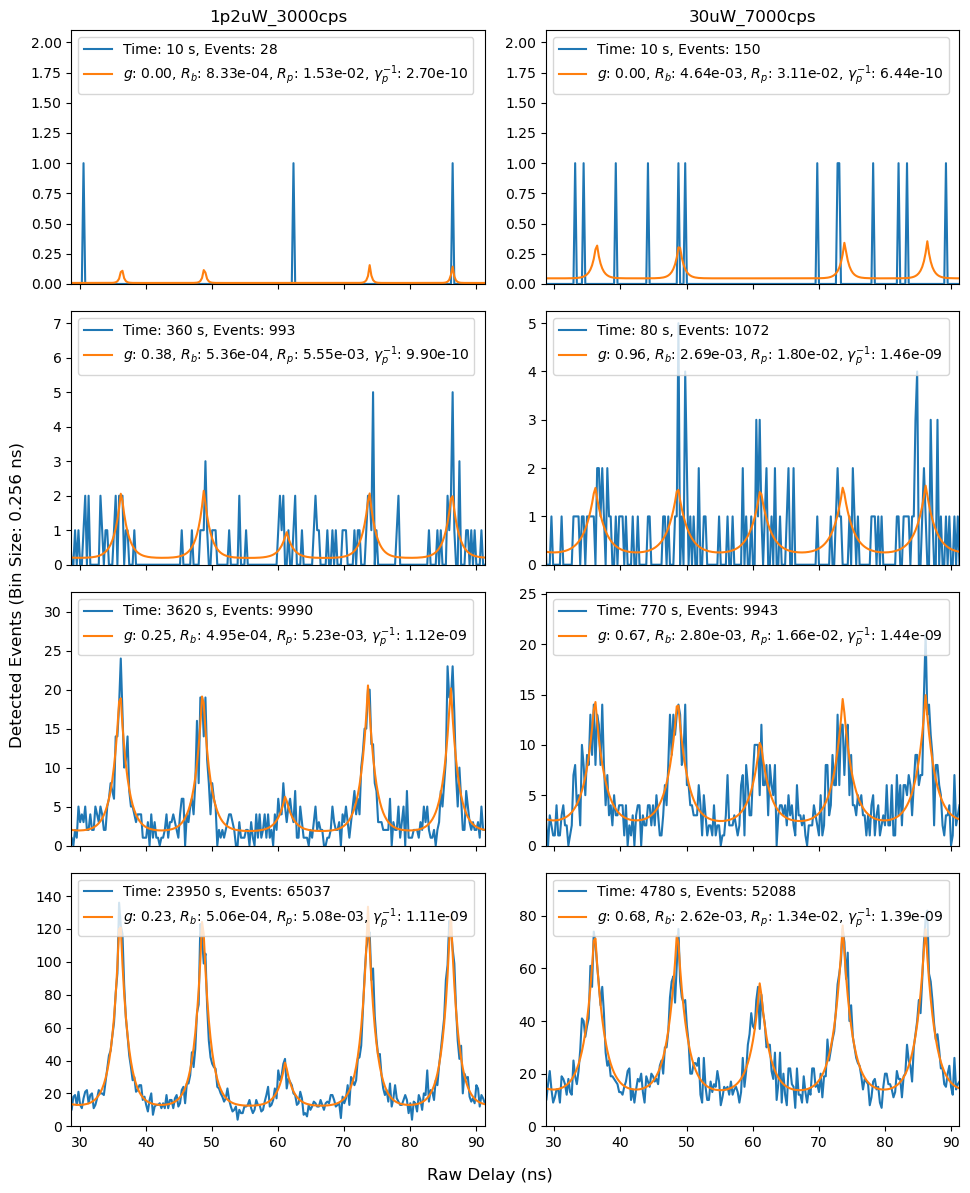}
\caption{\label{Fig:Hist} Example histograms for the 1p2uW (left column) and 30uW (right column) experimental contexts, where the raw time delay between triggered detectors denotes which bin of $0.256$ ns width a corresponding coincidence is assigned to. These are close-ups; the full $500$ ns domain contains $40$ peaks. Row 1: the first $10$ s snapshot of measurement. Row 2: the accumulation of ${\sim}1000$ detections. Row 3: the accumulation of ${\sim}10000$ detections. Row 4: the full duration of interferometry. Observations are overlaid with optimised fits of $R(\tau_i; \theta) \times t$. Units of $R_b$ and $R_p$ are Hz. Units of $1/\gamma_p$ are s. Estimate for $\tau_0$ quickly shifts from $61.4$ to $61.1$ ns between Row 1 and 4.}
\end{figure*}

Example histograms of the 1p2uW and 30uW experimental contexts are displayed in Fig.~\ref{Fig:Hist}; these represent HBT interferometry measurements of the SPS when subjected to the lowest and highest laser intensities of $1.2$ \si{\uW} and $30$ \si{\uW}.
As shown, the theoretical comb structure of Eq.~(\ref{Eq:Fit}) becomes much clearer as more coincidences are observed.
The monotonic relation between laser intensity and $g$ within the FI-SEQUR datasets is also exemplified by the MPE peak and its variation in relative size.
Additionally, the plots demonstrate that fixed numbers of detections, e.g.~$1000$ and $10000$, are accumulated over different measurement durations for different experimental contexts.

\subsection{The Estimators}
\label{Sec:Estimators}

\subsubsection{Standard Fitting}
\label{Sec:Fitting}

The conventional method of determining $g$ involves fitting a theoretically derived equation to observed data, i.e.~an accumulated histogram of two-photon event detections, which is labelled as $d_i$ and associated with time-delay bins indexed by $i$.
Here, as before~\cite{kemu23}, the parameters of Eq.~(\ref{Eq:Fit}) are optimised with the Python LMFIT package~\cite{nest14} whenever required:
\begin{equation}
\label{Eq:Objective}
    \theta_{LS} = \argmin_\theta \sum_i \left(R(\tau_i; \theta) - \frac{d_i}{t}\right)^2.
\end{equation}
Specifically, `least-squares' fitting is undertaken via the Powell method, which is paired with a subsequent Trust Region Reflective optimisation (\texttt{method="least\_squares"}).
Of course, there are other fitting methods and objective functions, some of which have been explored elsewhere~\cite{coad20}, but these require further benchmarking; least-squares fitting has not yet been convincingly superseded, at least for the FI-SEQUR data~\cite{kemu23}.
In any case, only five of the parameters in Eq.~(\ref{Eq:Fit}) are unknown for the FI-SEQUR datasets, as laser pulse period is a constant $\Lambda=1.25\times 10^{-8}$, while envelope decay is negligible, i.e.~$\gamma_e=0$.

\begin{table*}[htb!]
\centering
\caption{`Ground-truth' values with standard errors for the parameters of Eq.~(\ref{Eq:Fit}), as determined by least-squares fitting $R(\tau; \theta) \times t$ to all the data available in a FI-SEQUR dataset.}
\label{Tab:BestFit}
\begin{tabular}{|c|c|c|c|c|c|}
\hline
Best Fit  & $g$                & $\tau_0$ (s)                   & $R_b$ (Hz)                       & $R_p$ (Hz)                      & $1/\gamma_p$ (s)               \\ \hline
1p2uW  & 0.23 ($\pm 11.72\%$) & \num{6.11e-8} ($\pm 0.05\%$) & \num{2.54e-4} ($\pm 24.37\%$) & \num{2.55e-3} ($\pm 1.20\%$) & \num{1.10e-9} ($\pm 13.20\%$) \\
2p5uW  & 0.32 ($\pm 8.79\%$)  & \num{6.11e-8} ($\pm 0.06\%$) & \num{7.24e-4} ($\pm 39.84\%$) & \num{9.76e-3} ($\pm 1.37\%$) & \num{1.16e-9} ($\pm 15.13\%$) \\
4uW    & 0.36 ($\pm 7.31\%$)  & \num{6.11e-8} ($\pm 0.05\%$) & \num{9.13e-4} ($\pm 42.84\%$) & \num{1.30e-2} ($\pm 1.35\%$) & \num{1.18e-9} ($\pm 15.08\%$) \\
8uW    & 0.49 ($\pm 9.15\%$)  & \num{6.11e-8} ($\pm 0.10\%$) & \num{1.14e-3} ($\pm 79.85\%$) & \num{1.36e-2} ($\pm 2.75\%$) & \num{1.23e-9} ($\pm 32.09\%$) \\
10uW-- & 0.53 ($\pm 7.20\%$)  & \num{6.11e-8} ($\pm 0.08\%$) & \num{1.20e-3} ($\pm 91.99\%$) & \num{1.74e-2} ($\pm 2.54\%$) & \num{1.34e-9} ($\pm 27.54\%$) \\
10uW+  & 0.51 ($\pm 5.92\%$)  & \num{6.11e-8} ($\pm 0.07\%$) & \num{4.46e-3} ($\pm 78.09\%$) & \num{7.33e-2} ($\pm 1.93\%$) & \num{1.26e-9} ($\pm 22.14\%$) \\
20uW   & 0.56 ($\pm 5.94\%$)  & \num{6.11e-8} ($\pm 0.07\%$) & \num{2.18e-3} ($\pm 49.65\%$) & \num{1.78e-2} ($\pm 2.39\%$) & \num{1.34e-9} ($\pm 26.62\%$) \\
30uW   & 0.68 ($\pm 5.97\%$)  & \num{6.11e-8} ($\pm 0.09\%$) & \num{2.62e-3} ($\pm 50.86\%$) & \num{1.34e-2} ($\pm 3.68\%$) & \num{1.39e-9} ($\pm 41.80\%$) \\ \hline
\end{tabular}
\end{table*}

Because the observed detection rate, $d_i/t$, tends towards theoretical average rate $R_i$ when $t$ is sufficiently large, the uncertainty in $g$ and other fitted parameters gradually decreases for large enough sample sizes.
However, as shown previously~\cite{kemu23}, the uncertainty never completely vanishes even for long HBT interferometry sessions, meaning that ultra-low-$g$ experiments may need ultra-long-$t$ measurements to confidently prove an SPS is SOTA.
Nonetheless, fits applied to the final histograms of each FI-SEQUR dataset -- see the last row of Fig.~\ref{Fig:Hist} -- provide the best available parameter estimates per dataset, and so the values in Table~\ref{Tab:BestFit} are taken to be `ground truths'.
In fact, whenever this work requires new synthetic data representing a FI-SEQUR experimental context, it is the eight best-fit average-rate functions based on these ground-truth parameter values that are Poisson-sampled.
Details of this generative procedure are found elsewhere~\cite{kemu23}.

\subsubsection{Machine Learning}
\label{Sec:ML}

In a sense, ML can also be considered a form of fitting, mapping an input feature space to an output target space.
Specifically, this work explores ML predictors capable of representing a function $f: [0,1]^{1954} \rightarrow [0,1]$, i.e.~a mapping from $d_i/\sum_{i=1}^{1954}d_i$ to $g$.
The size of the input space is determined by the FI-SEQUR datasets, which, barring one, are originally presented in $1954$ time-delay bins.
Additionally, the normalisation of input space by the sum of all events detected, per histogram, is done to ensure that sample size does not confuse an ML predictor; its predictions should be based solely on the \textit{shape} of a histogram, given that $g$ is a relative ratio of the MPE peak to any other standard peak.

Now, given the right functional complexity, an ML model could potentially learn to mimic the least-squares fitting of $g$, i.e.~Eq.~(\ref{Eq:Objective}) applied to Eq.~(\ref{Eq:Fit}).
However, given that this form of least-squares fitting is already very efficient, this is not the appeal of ML in this work.
More interesting is whether an ML predictor can link patterns within $d_i$ to the \textit{eventual} estimate of $g$, i.e.~the ground truth for an experimental context.
Standard fitting, while ideal in the long run, is highly inaccurate in the short term~\cite{kemu23}.
The question is: can ML beat standard fitting by internalising existing knowledge of the $g$ value that a small-sample histogram will lead to?

There are a couple of challenges to face when answering such a question.
First, ML learns by differentiating between sets of both input feature vectors and output target values.
For instance, there is nothing informative to be learned from $2395$ normalised 1p2uW histograms alone if they all associate with a ground truth of $g=0.23$.
Any typical ML model faced with this training set would simply assume that \textit{any} histogram is from an experimental context with $g=0.23$.
Thus, ML models intended for predicting the best estimate of $g$ need to train on at least two datasets.

The other issue is more pragmatic.
When encountering a new context, the ground-truth $g$ value is unknown.
Even assuming that a from-scratch ML model can train on newly arriving data as quickly as possible, it has no apparent competitive advantage over standard fitting.
However, it may be possible to give ML this advantage by leveraging prior experience in other contexts, i.e.~by training an ML model on seven of eight FI-SEQUR datasets and then applying this knowledge to the `newly' encountered eighth dataset.
Here, this process is called `transfer learning', and the rest of this paper assesses whether it has any benefit compared to standard least-squares fitting.

In this work, five different ML predictors are investigated.
The most basic is linear regression via ordinary least squares (OLS), which leverages an analytical solution.
This solution has various optimality properties but can suffer from overfitting, i.e.~attributing too much importance to noise in the data.
Linear regression can also be trained iteratively via stochastic gradient descent (SGD), a more general and often faster procedure.
While SGD is iterative and may not converge to the OLS solution, this sometimes has a regularisation benefit, i.e.~it avoids overfitting.
Additionally, SGD is the only model explored here that can be adapted to new data.
Then, there is an alternative linear model, partial least squares regression (PLSR), which seeks to regularise by focussing on input-output covariance rather than correlation.

While the first three ML models are simple and almost free of hyperparameters, i.e.~settings users must specify to define model structure/training, the last two ML models are much more complex, capable of fitting nonlinearities, and are slow to train.
Random forest (RF) and gradient boosting (GB) are both ensemble models, here involving regression trees as individual learners, where the former takes the mean prediction of multiple trees, while the latter sequentially uses new learners to correct for the predictive errors of previous ones.
For all five ML models, linear or ensemble, this work also optionally pipelines in a standard scaler as a preprocessor, which, after the $d_i/\sum_{i=1}^{1954}d_i$ normalisation, scales the values across a training dataset of multiple histograms, per bin, to a mean of $0$ and a standard deviation of $1$.

During the following investigations, root mean squared error (RMSE) is used as the `loss' metric that describes fitting/model quality, i.e.~the average difference between fitting/model predictions and the actual values.
In fact, the RMSE is used even prior to assessing the transfer-learning quality of an ML model.
Indeed, given how many hyperparameter configurations are possible for a pipelined ML model, a presumptively optimal model must first be determined for the already encountered contexts via a process called hyperparameter optimisation (HPO).
In effect, numerous candidates, e.g.~SGD models with different learning rates, are trained and assessed on several splits of previously seen data.
The candidate with the lowest `validation loss', i.e.~an RMSE averaged across the splits, is considered to have the best model configuration, and this ML model is then freshly trained on \textit{all} the previous data, ready to be transferred to a new context and tested on its unseen data.

All ML processing in this work has been done with a yet-unpublished Python package, `AutonoML', inspired by a review of automated/autonomous ML~\cite{kemu20}.
However, for the purposes of this paper, the AutonoML codebase can be considered a convenience wrapper around other packages.
Indeed, all ML models in this work are pulled directly from Scikit-learn~\cite{peva11}, abbreviated as `sklearn', so readers are able to reproduce ML fitting/prediction on FI-SEQUR datasets themselves.
As for HPO, this is delegated to the HpBandSter package, which mixes bandit-based strategies with Bayesian optimisation~\cite{fakl18}.
For the sake of reproducibility, any HPO referred to in this work uses four iterations of successive `thirding' to narrow in on the best hyperparameter configuration, and the presumptive quality of a model is determined via 4-split Monte Carlo cross-validation applied to any relevant dataset, where each random split enforces a 3:1 ratio between training and validation data.

\section{Results}
\label{Sec:Results}

\subsection{Experimental Data Only}
\label{Sec:Experiment}

\begin{figure*}[p!]
\centering
\includegraphics[width=0.95\textwidth]{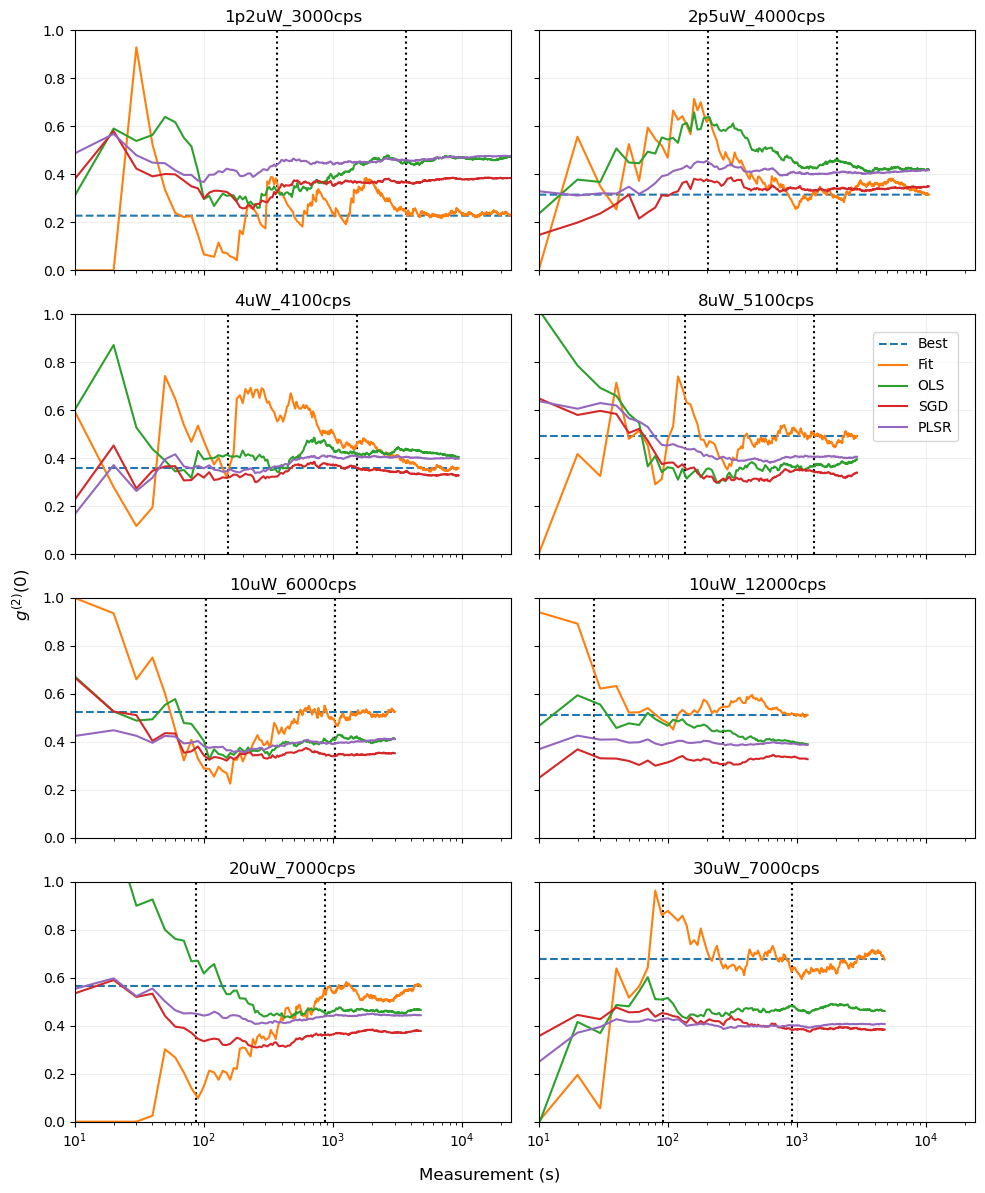}
\caption{\label{Fig:ExperimentLinear} Fitting and linear ML model estimates of $g$ for each FI-SEQUR dataset based on histograms of experimentally accumulated detections at every $10$ s timestep. The `Fit' trace denotes estimates made by least-squares fitting, with the final value defining the ground truth for each dataset, i.e.~a horizontal dashed `Best' line. The ML models, trained on the other seven contexts prior to estimation, are: ordinary least squares (OLS), stochastic gradient descent (SGD), and partial least squares regression (PLSR). The measurement-time axis is logarithmic, and the first and second vertical dotted lines mark the times at which $1000$ and $10000$ detections are accumulated, respectively, per dataset.}
\end{figure*}

This section limits SPS-quality analysis strictly to the real-world observations acquired by resource-expensive experiments.
For each FI-SEQUR context, i.e.~generally a different laser intensity, an accumulated histogram of coincidences is available at every $10$ seconds.
The function in Eq.~(\ref{Eq:Fit}) can be least-squares fitted to each histogram -- see Fig.~\ref{Fig:Hist} for examples -- and a series of $g$ estimates can be derived.
These `Fit' estimates are displayed in Fig.~\ref{Fig:ExperimentLinear}, and the last value obtained for each context becomes the `Best' measure of SPS quality, which is marked by dashed horizontal lines.

\begin{figure*}[p!]
\centering
\includegraphics[width=0.95\textwidth]{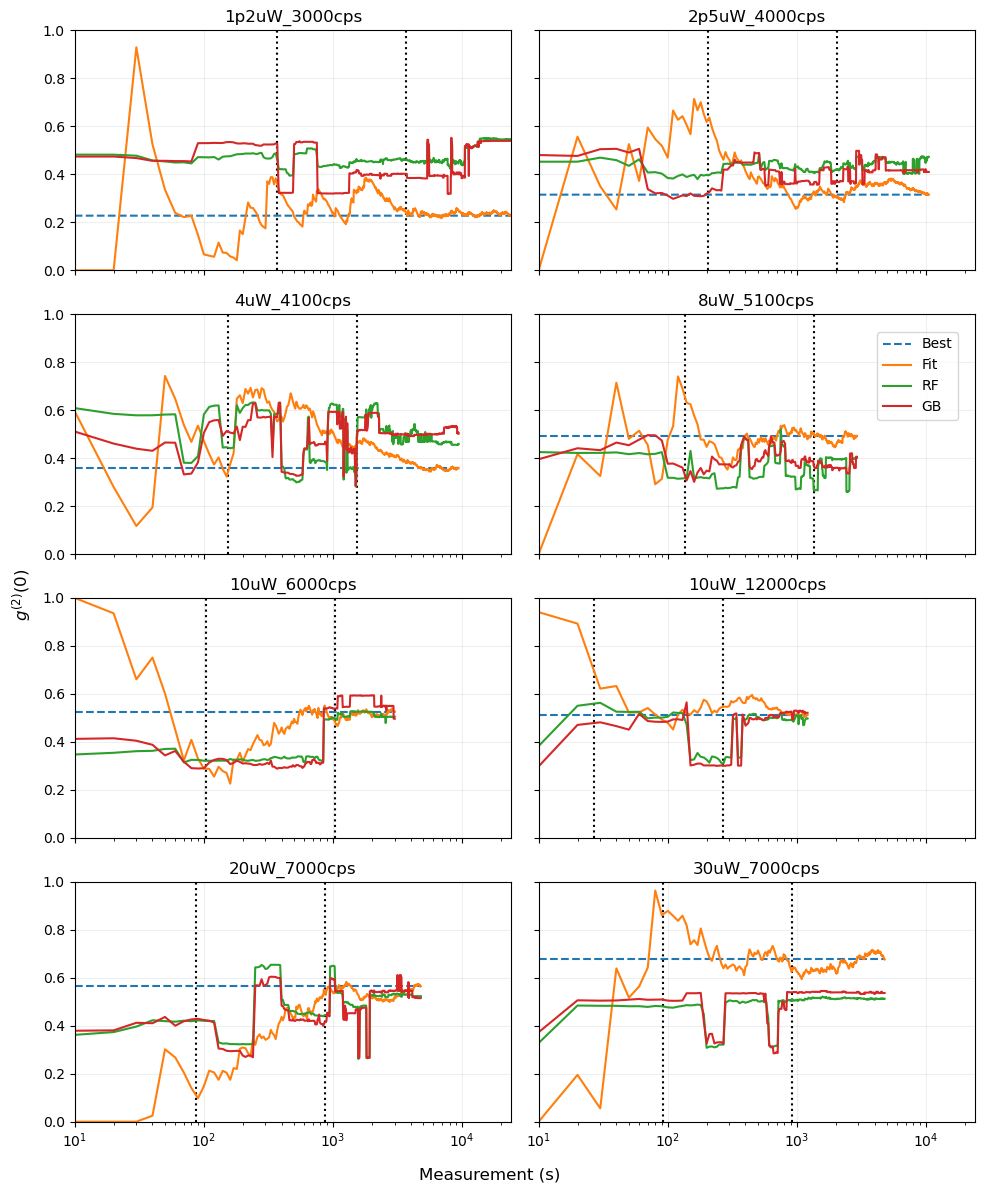}
\caption{\label{Fig:ExperimentEnsemble} Fitting and ensemble ML model estimates of $g$ for each FI-SEQUR dataset based on histograms of experimentally accumulated detections at every $10$ s timestep. The `Fit' trace denotes estimates made by least-squares fitting, with the final value defining the ground truth for each dataset, i.e.~a horizontal dashed `Best' line. The ML models, trained on the other seven contexts prior to estimation, are: random forest (RF) and gradient boosting (GB). The measurement-time axis is logarithmic, and the first and second vertical dotted lines mark the times at which $1000$ and $10000$ detections are accumulated, respectively, per dataset.}
\end{figure*}

Now, for each of the eight experimental contexts, ML models learn from the other seven.
Given that there are $6055$ cumulative histograms across all eight datasets, calculated by summing the total duration of experiments listed in Table~\ref{Tab:Datasets}, this means each model may train on $3660$ to $5934$ histograms of $1954$ bins and their associated `best' $g$ values.
Of course, described in Section~\ref{Sec:ML}, determining the optimal hyperparameter configuration of an ML model takes the bulk of computational time.
However, once an optimal ML model has been determined across the `previous' seven contexts, in terms of minimising validation loss, it is quick and simple to feed it the histograms of the `new' eighth context and generate forecasts of $g$.
Accordingly, the predictions of the linear ML models are contrasted against least-squares fitting in Fig.~\ref{Fig:ExperimentLinear}, while those of the ensemble ML models are displayed in Fig.~\ref{Fig:ExperimentEnsemble}.

\begin{table*}[htb!]
\centering
\caption{Optimal hyperparameters for ML models trained/tested solely on experimental data, as found by the HPO runs described in Section~\ref{Sec:ML}. Note that any sklearn model intended for a context was first trained/validated on the other seven contexts, and any hyperparameter not mentioned in this table is left as default. A `Y' in the Scaler column indicates training data is normalised to zero mean and unit variance, with test data rescaled appropriately. The learning rate and regularisation term for SGD refers to `eta0' and `alpha' in the sklearn package, respectively.}
\label{Tab:ExperimentHPO}
\begin{tabular}{|l|l|c|c|c|c|c|c|c|}
\toprule
Model & Context & Scaler & \makecell{Learning\\ Rate} & \makecell{Regularisation\\ Term} & Estimators & \makecell{Max\\ Depth} & \makecell{Min Samples\\ Per Leaf} & \makecell{Min Samples\\ To Split} \\
\midrule
\multirow{8}{*}{OLS} & 1p2uW & Y &  &  &  &  &  &  \\
 & 2p5uW & Y &  &  &  &  &  &  \\
 & 4uW & Y &  &  &  &  &  &  \\
 & 8uW & N &  &  &  &  &  &  \\
 & 10uW-- & N &  &  &  &  &  &  \\
 & 10uW+ & Y &  &  &  &  &  &  \\
 & 20uW & Y &  &  &  &  &  &  \\
 & 30uW & N &  &  &  &  &  &  \\
\cline{1-9}
\multirow{8}{*}{SGD} & 1p2uW & Y &  2.80e-04 &  1.27e-10 &  &  &  &  \\
 & 2p5uW & Y &  2.68e-04 &  2.56e-02 &  &  &  &  \\
 & 4uW & Y &  2.22e-04 &  1.37e-05 &  &  &  &  \\
 & 8uW & Y &  2.22e-04 &  4.98e-10 &  &  &  &  \\
 & 10uW-- & Y &  2.36e-04 &  7.05e-10 &  &  &  &  \\
 & 10uW+ & Y &  2.65e-04 &  3.39e-07 &  &  &  &  \\
 & 20uW & Y &  2.14e-04 &  2.33e-07 &  &  &  &  \\
 & 30uW & Y &  2.81e-04 &  2.98e-10 &  &  &  &  \\
\cline{1-9}
\multirow{8}{*}{PLSR} & 1p2uW & N &  &  &  &  &  &  \\
 & 2p5uW & N &  &  &  &  &  &  \\
 & 4uW & N &  &  &  &  &  &  \\
 & 8uW & Y &  &  &  &  &  &  \\
 & 10uW-- & Y &  &  &  &  &  &  \\
 & 10uW+ & N &  &  &  &  &  &  \\
 & 20uW & Y &  &  &  &  &  &  \\
 & 30uW & Y &  &  &  &  &  &  \\
\cline{1-9}
\multirow{8}{*}{RF} & 1p2uW & Y &  &  & 420 & 78 & 1 & 2 \\
 & 2p5uW & N &  &  & 131 & 34 & 1 & 2 \\
 & 4uW & N &  &  & 319 & 11 & 1 & 2 \\
 & 8uW & N &  &  & 157 & 23 & 1 & 2 \\
 & 10uW-- & Y &  &  & 344 & 26 & 1 & 4 \\
 & 10uW+ & N &  &  & 181 & 24 & 1 & 2 \\
 & 20uW & N &  &  & 143 & 47 & 1 & 3 \\
 & 30uW & Y &  &  & 170 & 33 & 1 & 2 \\
\cline{1-9}
\multirow{8}{*}{GB} & 1p2uW & N &  3.99e-02 &  & 188 & 23 & 3 & 32 \\
 & 2p5uW & Y &  8.68e-02 &  & 202 & 21 & 15 & 2 \\
 & 4uW & Y &  3.66e-02 &  & 182 & 3 & 2 & 13 \\
 & 8uW & Y &  8.91e-02 &  & 204 & 2 & 2 & 19 \\
 & 10uW-- & Y &  4.49e-02 &  & 160 & 29 & 16 & 20 \\
 & 10uW+ & N &  7.04e-02 &  & 92 & 32 & 1 & 13 \\
 & 20uW & Y &  1.22e-01 &  & 248 & 25 & 16 & 30 \\
 & 30uW & N &  8.65e-02 &  & 183 & 25 & 16 & 19 \\
\cline{1-9}
\bottomrule
\end{tabular}
\end{table*}

These initial results suggest that ML is not an automatic panacea to the problem of SPS quality estimation.
For instance, the long-term behaviour of the linear-model estimates appears bounded within the $0.3$ to $0.5$ range of $g$ values, meaning the predictions become pointedly inaccurate for the edge cases of low and high laser intensity.
This result is not unexpected; ML seeks to generalise its inductive understanding of data, and linear regression is the extreme of this representational simplicity, often failing to capture nuance and complexity within data via underfitting.
In contrast, ensemble models are designed to heavily partition input space, i.e.~the diversity of possible histograms, and capture possible distinctions and nonlinearities in the way different histograms associate with the best $g$ value.
Accordingly, their forecasts appear much more jerky as the cumulative histograms evolve, but, for example, they do align remarkably well in the long term for the $10$ and $20$ \si{\uW} contexts.
Unfortunately, the challenge with ensemble models, as Table~\ref{Tab:ExperimentHPO} shows, is that their hyperparameter spaces are large and thus difficult to optimise.
It is also possible that the large number of estimators, i.e.~trees, and their substantial depths, i.e.~the complexity of their regression rules, leads to overfitting, i.e.~seeing relationships where there are none.

\begin{table*}[htb!]
\centering
\caption{The loss values, i.e.~RMSE, for least-squares fitting and the transfer-learning ML models when tested on each context. Testing and (where applicable) training on the other seven contexts only involves experimental data. The prediction RMSE is calculated for histograms containing under $1000$ coincidences for `Early' loss, $1000$ to $10000$ coincidences for `Mid' loss, over $10000$ coincidences for `Late' loss, and any number of coincidences for `Full' loss. Loss is colour-coded: yellow to red for low to high values. Also included are the validation losses for ML models, i.e.~the RMSE when applied to the seven contexts they were trained on; see Section~\ref{Sec:ML}. Note that training and validation on these seven contexts use independent subsamples of data, i.e.~no model is validated on data it is trained on.}
\label{Tab:ExperimentLoss}
\begin{tabular}{llccccc}
\toprule
Context & Model & Loss (Early) & Loss (Mid) & Loss (Late) & Loss (Full) & Loss (Valid.) \\
\midrule\\[-3.5mm]
\multirow{6}{*}{1p2uW} & Fit & \cellcolor[rgb]{1.00,0.73,0.33} 0.1742 & \cellcolor[rgb]{1.00,0.91,0.59} 0.0769 & \cellcolor[rgb]{1.00,0.99,0.77} 0.0107 & \cellcolor[rgb]{1.00,0.96,0.70} 0.0371 &  \\
 & OLS & \cellcolor[rgb]{1.00,0.73,0.34} 0.1728 & \cellcolor[rgb]{0.99,0.65,0.28} 0.2080 & \cellcolor[rgb]{0.99,0.58,0.25} 0.2370 & \cellcolor[rgb]{0.99,0.60,0.25} 0.2323 & \cellcolor[rgb]{1.00,0.99,0.77} 0.0121 \\
 & SGD & \cellcolor[rgb]{1.00,0.87,0.50} 0.1127 & \cellcolor[rgb]{1.00,0.82,0.43} 0.1404 & \cellcolor[rgb]{1.00,0.78,0.39} 0.1538 & \cellcolor[rgb]{1.00,0.79,0.39} 0.1514 & \cellcolor[rgb]{1.00,0.88,0.52} 0.1025 \\
 & PLSR & \cellcolor[rgb]{1.00,0.68,0.29} 0.1950 & \cellcolor[rgb]{0.99,0.61,0.26} 0.2275 & \cellcolor[rgb]{0.99,0.56,0.24} 0.2448 & \cellcolor[rgb]{0.99,0.57,0.24} 0.2418 & \cellcolor[rgb]{1.00,0.96,0.69} 0.0406 \\
 & RF & \cellcolor[rgb]{0.99,0.56,0.24} 0.2470 & \cellcolor[rgb]{0.99,0.61,0.26} 0.2281 & \cellcolor[rgb]{0.99,0.41,0.19} 0.2868 & \cellcolor[rgb]{0.99,0.44,0.20} 0.2788 & \cellcolor[rgb]{1.00,0.99,0.78} 0.0073 \\
 & GB & \cellcolor[rgb]{0.99,0.41,0.19} 0.2863 & \cellcolor[rgb]{1.00,0.73,0.33} 0.1750 & \cellcolor[rgb]{0.99,0.48,0.21} 0.2689 & \cellcolor[rgb]{0.99,0.52,0.23} 0.2582 & \cellcolor[rgb]{1.00,1.00,0.79} 0.0057 \\
\cline{1-7}\\[-3.5mm]
\multirow{6}{*}{2p5uW} & Fit & \cellcolor[rgb]{0.99,0.45,0.21} 0.2755 & \cellcolor[rgb]{1.00,0.92,0.62} 0.0669 & \cellcolor[rgb]{1.00,0.96,0.70} 0.0356 & \cellcolor[rgb]{1.00,0.94,0.64} 0.0570 &  \\
 & OLS & \cellcolor[rgb]{0.99,0.60,0.25} 0.2317 & \cellcolor[rgb]{1.00,0.76,0.36} 0.1639 & \cellcolor[rgb]{1.00,0.87,0.51} 0.1094 & \cellcolor[rgb]{1.00,0.85,0.47} 0.1239 & \cellcolor[rgb]{1.00,0.98,0.76} 0.0171 \\
 & SGD & \cellcolor[rgb]{1.00,0.92,0.62} 0.0683 & \cellcolor[rgb]{1.00,0.97,0.72} 0.0283 & \cellcolor[rgb]{1.00,0.97,0.72} 0.0277 & \cellcolor[rgb]{1.00,0.97,0.72} 0.0291 & \cellcolor[rgb]{1.00,0.93,0.63} 0.0617 \\
 & PLSR & \cellcolor[rgb]{1.00,0.89,0.55} 0.0916 & \cellcolor[rgb]{1.00,0.89,0.55} 0.0937 & \cellcolor[rgb]{1.00,0.89,0.54} 0.0959 & \cellcolor[rgb]{1.00,0.89,0.54} 0.0955 & \cellcolor[rgb]{1.00,0.93,0.64} 0.0589 \\
 & RF & \cellcolor[rgb]{1.00,0.88,0.52} 0.1021 & \cellcolor[rgb]{1.00,0.86,0.49} 0.1153 & \cellcolor[rgb]{1.00,0.84,0.45} 0.1311 & \cellcolor[rgb]{1.00,0.84,0.46} 0.1280 & \cellcolor[rgb]{1.00,1.00,0.79} 0.0039 \\
 & GB & \cellcolor[rgb]{1.00,0.88,0.53} 0.0985 & \cellcolor[rgb]{1.00,0.90,0.57} 0.0855 & \cellcolor[rgb]{1.00,0.87,0.51} 0.1069 & \cellcolor[rgb]{1.00,0.88,0.52} 0.1033 & \cellcolor[rgb]{1.00,1.00,0.79} 0.0048 \\
\cline{1-7}\\[-3.5mm]
\multirow{6}{*}{4uW} & Fit & \cellcolor[rgb]{1.00,0.72,0.32} 0.1778 & \cellcolor[rgb]{1.00,0.69,0.29} 0.1919 & \cellcolor[rgb]{1.00,0.96,0.69} 0.0394 & \cellcolor[rgb]{1.00,0.90,0.57} 0.0850 &  \\
 & OLS & \cellcolor[rgb]{1.00,0.77,0.37} 0.1590 & \cellcolor[rgb]{1.00,0.91,0.60} 0.0744 & \cellcolor[rgb]{1.00,0.93,0.63} 0.0636 & \cellcolor[rgb]{1.00,0.92,0.62} 0.0679 & \cellcolor[rgb]{1.00,0.99,0.77} 0.0135 \\
 & SGD & \cellcolor[rgb]{1.00,0.94,0.64} 0.0569 & \cellcolor[rgb]{1.00,0.98,0.75} 0.0190 & \cellcolor[rgb]{1.00,0.97,0.74} 0.0237 & \cellcolor[rgb]{1.00,0.97,0.74} 0.0240 & \cellcolor[rgb]{1.00,0.92,0.60} 0.0738 \\
 & PLSR & \cellcolor[rgb]{1.00,0.93,0.64} 0.0595 & \cellcolor[rgb]{1.00,0.96,0.69} 0.0409 & \cellcolor[rgb]{1.00,0.95,0.68} 0.0431 & \cellcolor[rgb]{1.00,0.95,0.68} 0.0431 & \cellcolor[rgb]{1.00,0.93,0.64} 0.0597 \\
 & RF & \cellcolor[rgb]{1.00,0.68,0.29} 0.1970 & \cellcolor[rgb]{1.00,0.77,0.37} 0.1596 & \cellcolor[rgb]{1.00,0.83,0.44} 0.1357 & \cellcolor[rgb]{1.00,0.81,0.42} 0.1407 & \cellcolor[rgb]{1.00,0.99,0.78} 0.0071 \\
 & GB & \cellcolor[rgb]{1.00,0.84,0.45} 0.1294 & \cellcolor[rgb]{1.00,0.79,0.39} 0.1517 & \cellcolor[rgb]{1.00,0.78,0.38} 0.1560 & \cellcolor[rgb]{1.00,0.78,0.38} 0.1550 & \cellcolor[rgb]{1.00,0.99,0.78} 0.0068 \\
\cline{1-7}\\[-3.5mm]
\multirow{6}{*}{8uW} & Fit & \cellcolor[rgb]{1.00,0.68,0.29} 0.1967 & \cellcolor[rgb]{1.00,0.95,0.68} 0.0434 & \cellcolor[rgb]{1.00,0.98,0.75} 0.0194 & \cellcolor[rgb]{1.00,0.94,0.66} 0.0521 &  \\
 & OLS & \cellcolor[rgb]{0.99,0.65,0.28} 0.2104 & \cellcolor[rgb]{1.00,0.82,0.43} 0.1375 & \cellcolor[rgb]{1.00,0.86,0.49} 0.1169 & \cellcolor[rgb]{1.00,0.84,0.45} 0.1312 & \cellcolor[rgb]{1.00,0.99,0.77} 0.0115 \\
 & SGD & \cellcolor[rgb]{1.00,0.88,0.53} 0.0993 & \cellcolor[rgb]{1.00,0.76,0.36} 0.1622 & \cellcolor[rgb]{1.00,0.76,0.37} 0.1615 & \cellcolor[rgb]{1.00,0.77,0.37} 0.1595 & \cellcolor[rgb]{1.00,0.91,0.59} 0.0792 \\
 & PLSR & \cellcolor[rgb]{1.00,0.91,0.58} 0.0818 & \cellcolor[rgb]{1.00,0.89,0.55} 0.0928 & \cellcolor[rgb]{1.00,0.90,0.56} 0.0887 & \cellcolor[rgb]{1.00,0.89,0.55} 0.0901 & \cellcolor[rgb]{1.00,0.94,0.64} 0.0575 \\
 & RF & \cellcolor[rgb]{1.00,0.87,0.49} 0.1149 & \cellcolor[rgb]{1.00,0.77,0.38} 0.1581 & \cellcolor[rgb]{1.00,0.84,0.45} 0.1319 & \cellcolor[rgb]{1.00,0.81,0.41} 0.1428 & \cellcolor[rgb]{1.00,0.99,0.77} 0.0098 \\
 & GB & \cellcolor[rgb]{1.00,0.91,0.59} 0.0778 & \cellcolor[rgb]{1.00,0.88,0.52} 0.1034 & \cellcolor[rgb]{1.00,0.84,0.46} 0.1277 & \cellcolor[rgb]{1.00,0.86,0.49} 0.1162 & \cellcolor[rgb]{1.00,0.99,0.78} 0.0084 \\
\cline{1-7}\\[-3.5mm]
\multirow{6}{*}{10uW--} & Fit & \cellcolor[rgb]{0.99,0.56,0.24} 0.2495 & \cellcolor[rgb]{1.00,0.88,0.53} 0.0978 & \cellcolor[rgb]{1.00,0.98,0.76} 0.0172 & \cellcolor[rgb]{1.00,0.92,0.60} 0.0725 &  \\
 & OLS & \cellcolor[rgb]{1.00,0.92,0.60} 0.0741 & \cellcolor[rgb]{1.00,0.82,0.43} 0.1401 & \cellcolor[rgb]{1.00,0.86,0.49} 0.1164 & \cellcolor[rgb]{1.00,0.85,0.47} 0.1232 & \cellcolor[rgb]{1.00,0.99,0.77} 0.0121 \\
 & SGD & \cellcolor[rgb]{1.00,0.84,0.46} 0.1274 & \cellcolor[rgb]{1.00,0.73,0.33} 0.1765 & \cellcolor[rgb]{1.00,0.73,0.33} 0.1762 & \cellcolor[rgb]{1.00,0.73,0.33} 0.1749 & \cellcolor[rgb]{1.00,0.91,0.60} 0.0748 \\
 & PLSR & \cellcolor[rgb]{1.00,0.86,0.49} 0.1156 & \cellcolor[rgb]{1.00,0.82,0.43} 0.1373 & \cellcolor[rgb]{1.00,0.86,0.48} 0.1200 & \cellcolor[rgb]{1.00,0.85,0.46} 0.1255 & \cellcolor[rgb]{1.00,0.94,0.65} 0.0563 \\
 & RF & \cellcolor[rgb]{1.00,0.72,0.32} 0.1815 & \cellcolor[rgb]{1.00,0.73,0.33} 0.1742 & \cellcolor[rgb]{1.00,0.98,0.76} 0.0173 & \cellcolor[rgb]{1.00,0.88,0.52} 0.1038 & \cellcolor[rgb]{1.00,0.99,0.78} 0.0083 \\
 & GB & \cellcolor[rgb]{1.00,0.72,0.32} 0.1814 & \cellcolor[rgb]{1.00,0.68,0.29} 0.1942 & \cellcolor[rgb]{1.00,0.95,0.67} 0.0473 & \cellcolor[rgb]{1.00,0.86,0.48} 0.1197 & \cellcolor[rgb]{1.00,0.99,0.78} 0.0073 \\
\cline{1-7}\\[-3.5mm]
\multirow{6}{*}{10uW+} & Fit & \cellcolor[rgb]{0.82,0.05,0.13} 0.4052 & \cellcolor[rgb]{1.00,0.95,0.68} 0.0434 & \cellcolor[rgb]{1.00,0.96,0.70} 0.0366 & \cellcolor[rgb]{1.00,0.93,0.63} 0.0643 &  \\
 & OLS & \cellcolor[rgb]{1.00,0.92,0.62} 0.0667 & \cellcolor[rgb]{1.00,0.95,0.67} 0.0470 & \cellcolor[rgb]{1.00,0.87,0.51} 0.1060 & \cellcolor[rgb]{1.00,0.89,0.54} 0.0966 & \cellcolor[rgb]{1.00,0.99,0.77} 0.0134 \\
 & SGD & \cellcolor[rgb]{0.99,0.64,0.27} 0.2121 & \cellcolor[rgb]{1.00,0.69,0.29} 0.1920 & \cellcolor[rgb]{1.00,0.71,0.31} 0.1817 & \cellcolor[rgb]{1.00,0.71,0.31} 0.1843 & \cellcolor[rgb]{1.00,0.94,0.66} 0.0511 \\
 & PLSR & \cellcolor[rgb]{1.00,0.86,0.48} 0.1182 & \cellcolor[rgb]{1.00,0.87,0.49} 0.1136 & \cellcolor[rgb]{1.00,0.86,0.48} 0.1210 & \cellcolor[rgb]{1.00,0.86,0.48} 0.1195 & \cellcolor[rgb]{1.00,0.94,0.65} 0.0555 \\
 & RF & \cellcolor[rgb]{1.00,0.89,0.54} 0.0956 & \cellcolor[rgb]{1.00,0.84,0.45} 0.1298 & \cellcolor[rgb]{1.00,0.94,0.65} 0.0549 & \cellcolor[rgb]{1.00,0.91,0.59} 0.0765 & \cellcolor[rgb]{1.00,1.00,0.79} 0.0042 \\
 & GB & \cellcolor[rgb]{1.00,0.78,0.38} 0.1544 & \cellcolor[rgb]{1.00,0.79,0.40} 0.1498 & \cellcolor[rgb]{1.00,0.93,0.63} 0.0627 & \cellcolor[rgb]{1.00,0.90,0.56} 0.0891 & \cellcolor[rgb]{1.00,1.00,0.79} 0.0036 \\
\cline{1-7}\\[-3.5mm]
\multirow{6}{*}{20uW} & Fit & \cellcolor[rgb]{0.64,0.00,0.15} 0.4632 & \cellcolor[rgb]{1.00,0.69,0.29} 0.1931 & \cellcolor[rgb]{1.00,0.96,0.70} 0.0364 & \cellcolor[rgb]{1.00,0.88,0.52} 0.1038 &  \\
 & OLS & \cellcolor[rgb]{0.89,0.10,0.11} 0.3733 & \cellcolor[rgb]{1.00,0.88,0.52} 0.1031 & \cellcolor[rgb]{1.00,0.88,0.52} 0.1036 & \cellcolor[rgb]{1.00,0.87,0.49} 0.1135 & \cellcolor[rgb]{1.00,0.99,0.77} 0.0099 \\
 & SGD & \cellcolor[rgb]{1.00,0.86,0.48} 0.1207 & \cellcolor[rgb]{0.99,0.61,0.26} 0.2268 & \cellcolor[rgb]{1.00,0.69,0.30} 0.1910 & \cellcolor[rgb]{1.00,0.68,0.29} 0.1963 & \cellcolor[rgb]{1.00,0.89,0.55} 0.0931 \\
 & PLSR & \cellcolor[rgb]{1.00,0.92,0.60} 0.0732 & \cellcolor[rgb]{1.00,0.82,0.43} 0.1394 & \cellcolor[rgb]{1.00,0.86,0.48} 0.1210 & \cellcolor[rgb]{1.00,0.85,0.47} 0.1236 & \cellcolor[rgb]{1.00,0.94,0.65} 0.0556 \\
 & RF & \cellcolor[rgb]{1.00,0.76,0.36} 0.1628 & \cellcolor[rgb]{1.00,0.82,0.43} 0.1372 & \cellcolor[rgb]{1.00,0.91,0.59} 0.0767 & \cellcolor[rgb]{1.00,0.89,0.55} 0.0913 & \cellcolor[rgb]{1.00,0.99,0.78} 0.0079 \\
 & GB & \cellcolor[rgb]{1.00,0.77,0.38} 0.1582 & \cellcolor[rgb]{1.00,0.77,0.38} 0.1580 & \cellcolor[rgb]{1.00,0.91,0.59} 0.0785 & \cellcolor[rgb]{1.00,0.88,0.53} 0.0977 & \cellcolor[rgb]{1.00,0.99,0.78} 0.0070 \\
\cline{1-7}\\[-3.5mm]
\multirow{6}{*}{30uW} & Fit & \cellcolor[rgb]{0.90,0.12,0.11} 0.3697 & \cellcolor[rgb]{1.00,0.94,0.65} 0.0553 & \cellcolor[rgb]{1.00,0.96,0.71} 0.0321 & \cellcolor[rgb]{1.00,0.93,0.63} 0.0627 &  \\
 & OLS & \cellcolor[rgb]{0.99,0.37,0.18} 0.2961 & \cellcolor[rgb]{0.99,0.64,0.27} 0.2129 & \cellcolor[rgb]{1.00,0.66,0.28} 0.2049 & \cellcolor[rgb]{0.99,0.65,0.28} 0.2084 & \cellcolor[rgb]{1.00,0.99,0.78} 0.0083 \\
 & SGD & \cellcolor[rgb]{0.99,0.59,0.25} 0.2363 & \cellcolor[rgb]{0.99,0.47,0.21} 0.2700 & \cellcolor[rgb]{0.99,0.39,0.19} 0.2898 & \cellcolor[rgb]{0.99,0.41,0.19} 0.2856 & \cellcolor[rgb]{1.00,0.89,0.54} 0.0949 \\
 & PLSR & \cellcolor[rgb]{0.99,0.41,0.19} 0.2870 & \cellcolor[rgb]{0.99,0.45,0.21} 0.2764 & \cellcolor[rgb]{0.99,0.47,0.21} 0.2712 & \cellcolor[rgb]{0.99,0.46,0.21} 0.2724 & \cellcolor[rgb]{1.00,0.94,0.65} 0.0540 \\
 & RF & \cellcolor[rgb]{0.99,0.63,0.27} 0.2168 & \cellcolor[rgb]{0.99,0.59,0.25} 0.2346 & \cellcolor[rgb]{1.00,0.76,0.36} 0.1631 & \cellcolor[rgb]{1.00,0.72,0.32} 0.1785 & \cellcolor[rgb]{1.00,0.99,0.78} 0.0067 \\
 & GB & \cellcolor[rgb]{1.00,0.70,0.30} 0.1882 & \cellcolor[rgb]{0.99,0.61,0.26} 0.2260 & \cellcolor[rgb]{1.00,0.82,0.43} 0.1377 & \cellcolor[rgb]{1.00,0.77,0.38} 0.1574 & \cellcolor[rgb]{1.00,1.00,0.79} 0.0053 \\
\cline{1-7}
\bottomrule
\end{tabular}
\end{table*}

Thus, based on the long-term results, one could argue that none of the ML models explored in this work compete with least-squares fitting.
Is ML simply not suited for this problem?
Well, Table~\ref{Tab:ExperimentLoss} suggests ML is actually very powerful.
On average, across the eight experimental contexts, least-squares fitting suggests $g$ values that are $0.0668$ off its eventual best estimate.
Even averaging the `late' loss instead of the `full' loss, i.e.~for stably shaped histograms containing over $10000$ coincidences, fitting is still $0.0284$ off the `ground truth'.
Compare then the validation losses of the ML models, which are essentially their inaccuracies when trained on one subset of seven experimental contexts and tested on another independent subset of those same seven contexts.
It is clear that OLS and the ensemble models have \textit{consistently} better predictive power than simply fitting Eq.~(\ref{Eq:Fit}), with their worst validation loss being $0.0171$.
Recalling from Table~\ref{Tab:BestFit} that a random sample of data from seven experimental contexts will consist of at least several target $g$ values that are $0.33$ apart, this accuracy is not coincidental.
Even PLSR and SGD are respectable competitors for least-squares fitting when trained on the same context.
So, the fact that ML models struggle to compete with least-squares fitting overall when applied to a \textit{new} context is more a question for the suitability of \textit{transfer} learning.
How genuinely similar are these contexts?

\subsection{Synthetic Data}
\label{Sec:Synthetic}

At this point, the powerful potential of ML to tease out hidden associations between detection histograms and SPS quality, beyond the standard process of fitting Eq.~(\ref{Eq:Fit}), has been shown.
Unfortunately, if a new SPS context is sufficiently different, the prior experience of an ML model may prove useless or even detrimental, due to its bias, compared to agnostic fitting.
This concept of context `difference' requires a broader discussion.

Nonetheless, what is clear from Fig.~\ref{Fig:ExperimentLinear} and Fig.~\ref{Fig:ExperimentEnsemble} is that early estimates provided by least-squares fitting are often highly unreliable.
Indeed, inspection of the `early' loss in Table~\ref{Tab:ExperimentLoss}, with RMSE values calculated for histograms containing under $1000$ coincidences, suggests ML predictions appear almost always superior, particularly when excluding the 1p2uW context.
Some caution with these results is required -- there are only two `early' histograms for the 10uW+ context -- but they do prompt the question: is transfer learning statistically better than least-squares fitting?

Experimental data is insufficient to tackle such a question, so this investigation leverages previous work~\cite{kemu23} to synthesise histograms that contain $1000$, $10000$, $100000$ and $1000000$ coincidences.
Note that none of the eight HBT interferometry experiments actually accumulate more than ${\sim}65000$ detections; the extreme late-stage histograms are synthesised to explore whether their minute variability is at all informative.
Crucially, each synthetic histogram is generated so that, in terms of Poisson statistics~\cite{coad20}, they are realistic to encounter within one of the eight SPS contexts described by the ground-truth parameters in Table~\ref{Tab:BestFit}.
Put another way, for each vertical dotted line in Fig.~\ref{Fig:ExperimentLinear}, and another two beyond the end of the plots, there are now $2500$ more histograms associated with a ground-truth $g$ to work with.
So, the analysis in this section begins by investigating the performance of transfer-learning models trained on seven contexts and tested on the eighth, but, for now, the training and testing involves \textit{only} histograms of a certain size, e.g.~$1000$, $10000$, $100000$ or $1000000$.
This means that the ML models, both in learning and application, are much more tightly focussed on a particular noise level for histogram shapes; compare row two and row three of Fig.~\ref{Fig:Hist}.
Optimised hyperparameters are omitted for considerations of space.

\begin{figure*}[p!]
\centering
\includegraphics[width=0.95\textwidth]{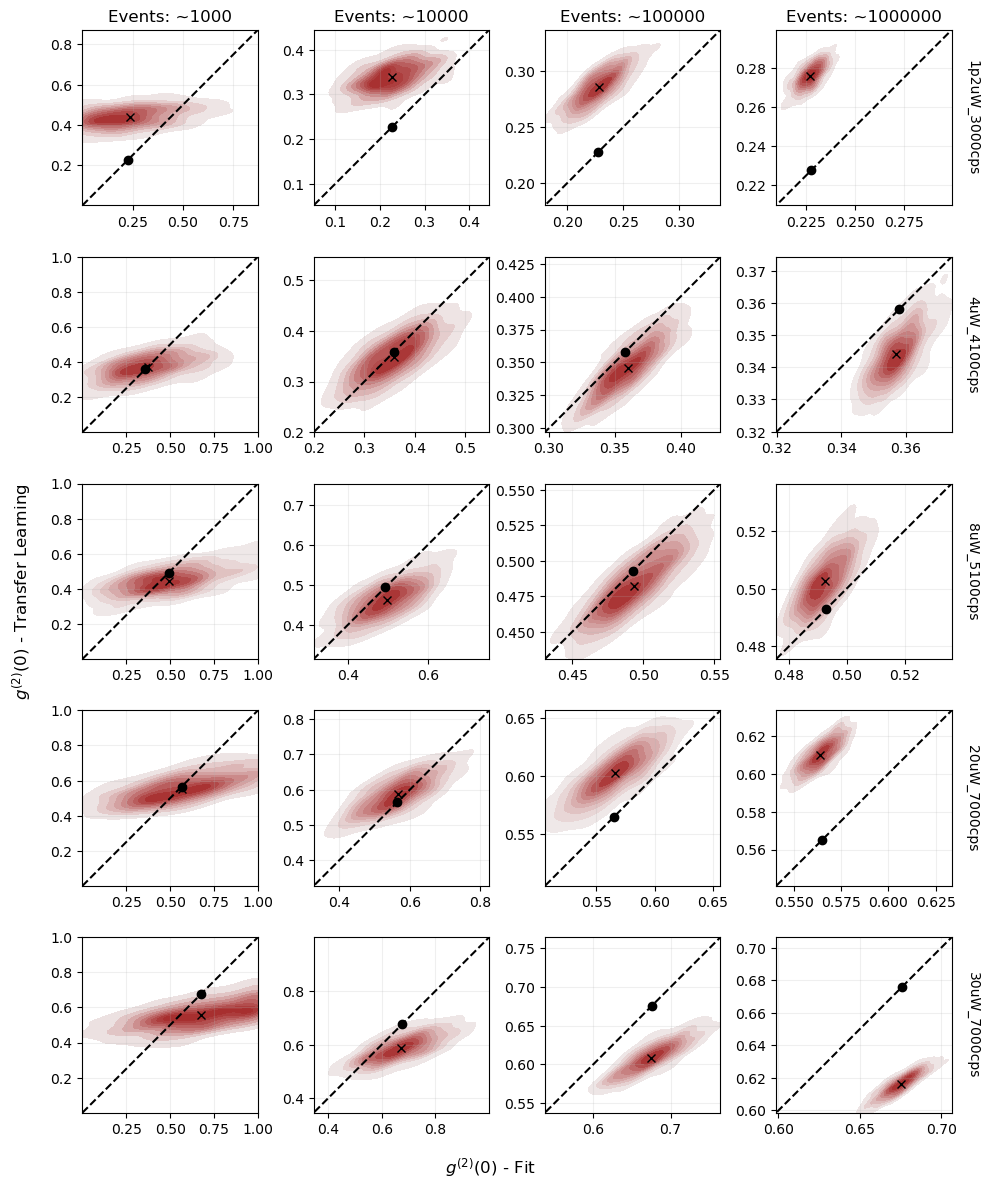}
\caption{\label{Fig:SynthOLS} Density plots of estimate comparisons between OLS models and least-squares fitting. Each red density `cloud' represents $500$ context-specific histograms containing a certain number of coincidences, e.g.~$1000$ detections for the 1p2uW context. For each histogram, the least-squares fit suggests a $g$ value along the $x$ axis and the transferred OLS model suggests a $g$ value along the $y$ axis. The dashed diagonal indicates where both estimates would be the same, and the black circle denotes the ground truth. The `x' marks the average prediction of both approaches. Per plot, axis scales are equal. For considerations of space, only five contexts are shown. Note that if the OLS model is tested on $500$ size-$1000000$ histograms, it was trained on $2000\times7$ size-$1000000$ histograms from seven other contexts.}
\end{figure*}

Consequently, the density plots in Fig.~\ref{Fig:SynthOLS} show how OLS models trained on fixed-size histograms from seven contexts perform when applied to histograms of the same fixed size in the eight context.
In greater detail, per context and histogram size, both least-squares fitting and OLS models make predictions of $g$ for $500$ input histograms, but the ML models are first trained on $2000 \times 7$ histograms from the other contexts, which is far more data than is experimentally available.
Upon examining the paired predictions, it is clear that, at least for the $g$ values discussed in this work, the estimates of the least-squares fit always average to align with the ground truth, i.e.~the `x' is always vertically aligned with the circle in the figure.
However, the stochastic variability in coincidence detection makes the shape of early histograms very noisy and thus induces a large variance in fitting estimates.
In contrast, while OLS often displays a systematic bias, presumably acquired from training on the other seven contexts, its ability to generalise makes its distribution of $g$ estimates much tighter and robust against noise.
Thus, provided an ML model is not operating within a context it is apparently not equipped to deal with, e.g.~1p2uW, it can still be a preferable \textit{early} estimator, statistically speaking.

\begin{figure*}[p!]
\centering
\includegraphics[width=0.95\textwidth]{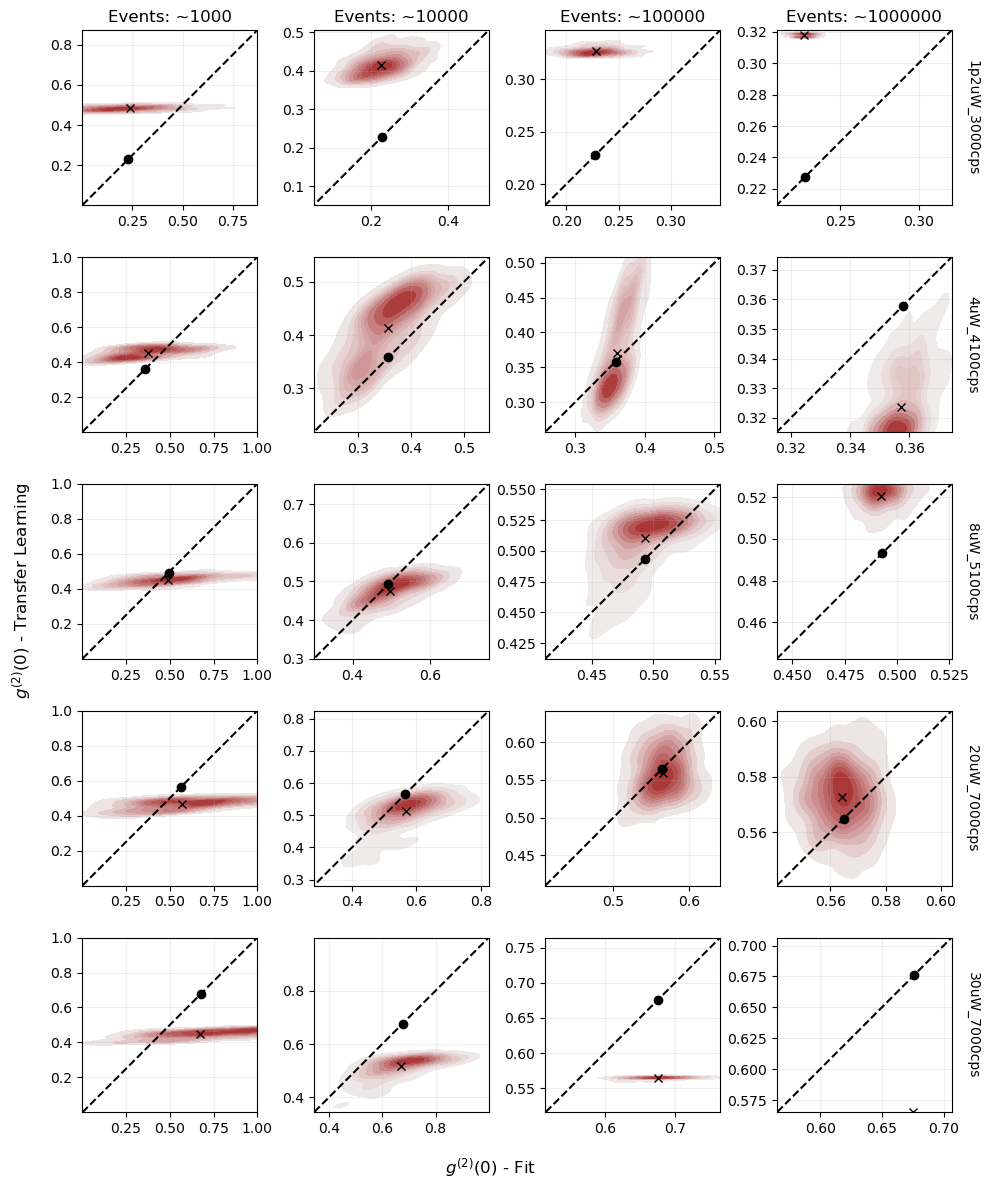}
\caption{\label{Fig:SynthRF} Density plots of estimate comparisons between RF models and least-squares fitting. Each red density `cloud' represents $500$ context-specific histograms containing a certain number of coincidences, e.g.~$1000$ detections for the 1p2uW context. For each histogram, the least-squares fit suggests a $g$ value along the $x$ axis and the transferred RF model suggests a $g$ value along the $y$ axis. The dashed diagonal indicates where both estimates would be the same, and the black circle denotes the ground truth. The `x' marks the average prediction of both approaches. Per plot, axis scales are equal. For considerations of space, only five contexts are shown. Note that if the RF model is tested on $500$ size-$1000000$ histograms, it was trained on $400\times7$ size-$1000000$ histograms from seven other contexts.}
\end{figure*}

The comparative synthetic-data results for SGD look similar to those for OLS.
However, the density plots for the ensemble ML models appear somewhat different, with those for RF displayed in Fig.~\ref{Fig:SynthRF}.
Unlike for the linear models, which were trained on $2000 \times 7$ histograms with ease, the lengthy HPO process for ensemble models necessitated a smaller training set of $400 \times 7$ data points.
Regardless, both PLSR and the ensemble models make relatively consistent $g$ predictions when trained and tested on size-$1000$ histograms, exhibiting far less variance than OLS/SGD and least-squares fitting; see the first column in the figure.
For PLSR, this relatively tight variance persists irrespective of training/testing histogram size, although the bias often becomes more and more noticeable for larger sizes, similar to the top and bottom rows of the RF prediction plots in Fig.~\ref{Fig:SynthRF}.
In contrast, the ensemble models are more complex in their predictive response, sometimes averaging quite close to the ground truth; see the central three plots of the third column in the figure.

\begin{figure*}[htb!]
\centering
\includegraphics[width=0.9\textwidth]{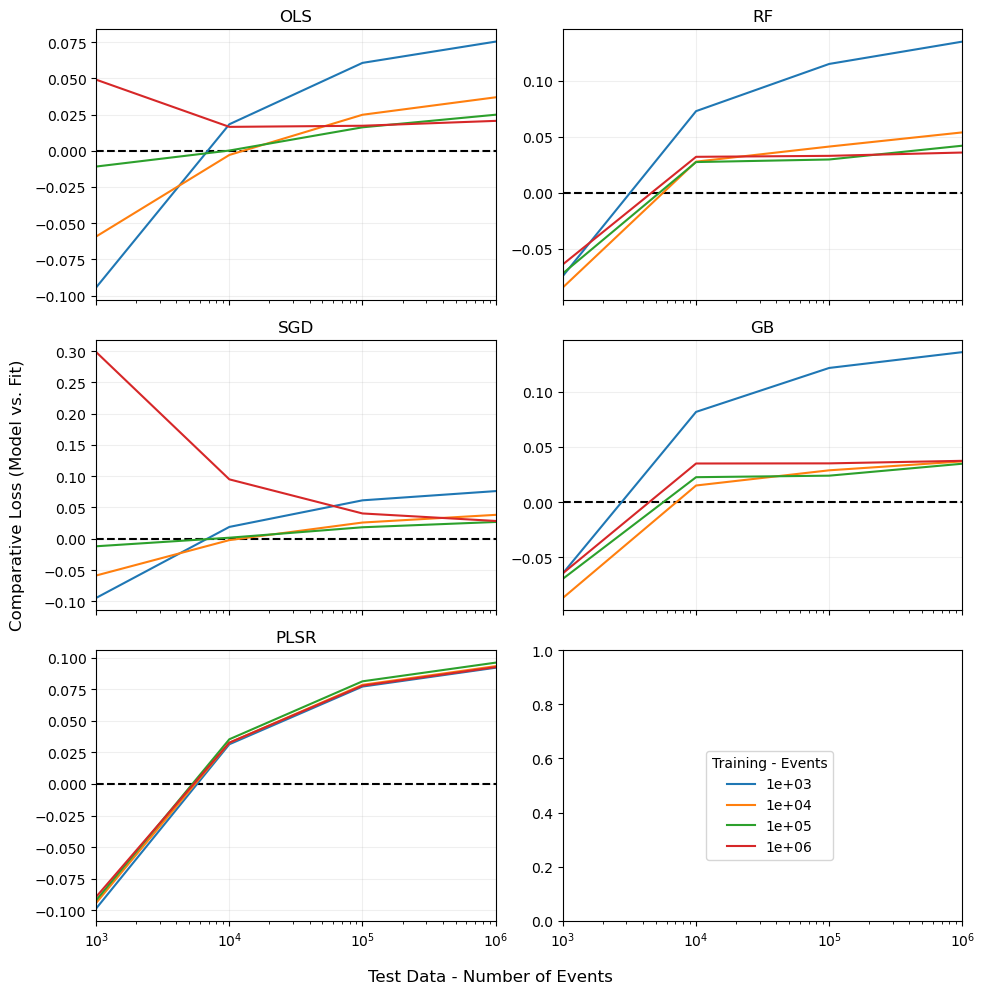}
\caption{\label{Fig:SynthAverage} The comparative loss of transfer-learning ML models versus the least-squares fitting of Eq.~(\ref{Eq:Fit}), i.e.~transfer-learning RMSE minus fitting RMSE. The losses are averaged across all eight contexts, and any comparative loss under $0$ suggests transfer-learning predictions are statistically more accurate. Per plot, each of the four lines represents an ML model being trained on a fixed histogram size, and the four points per line denote the histogram sizes that the model and fitting procedure are tested on. OLS, SGD and PLSR are trained on $14000$ histograms, while RF and GB are trained on $2800$.}
\end{figure*}

Ultimately, the first and second columns of Fig.~\ref{Fig:SynthOLS} and Fig.~\ref{Fig:SynthRF} are most important to the SPS use case, as they represent early estimation.
Nonetheless, the third and fourth column still shed insight on how ML models train on smooth histograms.
After all, a question lingers: could late-stage histograms help ML models predict $g$ for early-stage histograms?
For OLS and SGD, the answer is emphatically no, as shown by comparative-loss plots in Fig.~\ref{Fig:SynthAverage}.
The OLS/SGD models that do best when estimating $g$ from size-$n$ histograms are those \textit{trained} on size-$n$ histograms.
Such a `like-for-like' rule does not seem as evident for the ensemble models, primarily because their complexity seems to struggle and possibly overfit when faced with the noise of the size-$1000$ histograms.
A full training set of $14000$ histograms rather than $2800$ could also have an effect, but this is not feasible to explore with presently available computational hardware.

What is clear from Fig.~\ref{Fig:SynthAverage} is that least-squares fitting quickly becomes statistically more accurate than any transferred ML model studied here, so the benefits of ML are most apparent when only a few detections have been accumulated, i.e.~${\sim}1000$.
For OLS/SGD, the models are best when trained on similarly noisy samples, whereas the evidence is less clear for PLSR and RF/GB.
In the case of PLSR, its focus on covariance between histogram inputs and ground-truth $g$ outputs may make this form of ML too regularised for any nuance, thus making the model insensitive to what it is trained on.
In contrast, RF/GB might be complex enough to identify and process variability in the relatively smooth size-$1000000$ histograms, allowing useful associative rules to be transferred to the size-$1000$ case.
More research is required to untangle such concepts.

\subsection{Adaptive Transfer Learning}
\label{Sec:Adaptive}

In the ML sub-field of deep learning \cite{doke21}, transfer learning often involves not just taking a model from one context to another but also tweaking it somehow, e.g.~adjusting neural connection weights at the final predictive layers of a neural network.
Similarly, although this initial research into ML and the early estimation of SPS quality does not involve complex neural networks, it is worth exploring whether a transferred model can still be adapted to information from a new context.
Of course, adaptation is itself a very expansive topic covering many principles and methodologies \cite{zlbi12,baga18}, so this section only presents a rudimentary investigation in this space.

\begin{table*}[htb!]
\centering
\caption{Optimal hyperparameters for SGD models trained either solely on $3660$--$5934$ histograms of experimental data or solely on $14000$ synthetic histograms of size $1000$, as found by the HPO runs described in Section~\ref{Sec:ML}. Note that any sklearn model intended for a context was first trained/validated on the other seven contexts, and any hyperparameter not mentioned in this table is left as default. A `Y' in the Scaler column indicates training data is normalised to zero mean and unit variance, with test data rescaled appropriately. The learning rate and regularisation term for SGD refers to `eta0' and `alpha' in the sklearn package, respectively.}
\label{Tab:AdaptiveHPO}
\begin{tabular}{|l|l|c|c|c|}
\toprule
Model & Context & Scaler & \makecell{Learning\\ Rate} & \makecell{Regularisation\\ Term} \\
\midrule
\multirow{8}{*}{SGD-Exp} & 1p2uW & Y &  2.80e-04 &  1.27e-10 \\
 & 2p5uW & Y &  2.68e-04 &  2.56e-02 \\
 & 4uW & Y &  2.22e-04 &  1.37e-05 \\
 & 8uW & Y &  2.22e-04 &  4.98e-10 \\
 & 10uW-- & Y &  2.36e-04 &  7.05e-10 \\
 & 10uW+ & Y &  2.65e-04 &  3.39e-07 \\
 & 20uW & Y &  2.14e-04 &  2.33e-07 \\
 & 30uW & Y &  2.81e-04 &  2.98e-10 \\
\cline{1-5}
\multirow{8}{*}{SGD-Syn} & 1p2uW & Y &  1.34e-03 &  2.37e-07 \\
 & 2p5uW & Y &  9.87e-04 &  7.28e-06 \\
 & 4uW & Y &  6.65e-04 &  2.45e-08 \\
 & 8uW & Y &  6.22e-04 &  1.07e-02 \\
 & 10uW-- & Y &  1.20e-03 &  4.53e-02 \\
 & 10uW+ & Y &  7.77e-04 &  2.50e-07 \\
 & 20uW & Y &  2.16e-03 &  9.38e-02 \\
 & 30uW & Y &  3.88e-03 &  7.91e-02 \\
\cline{1-5}
\bottomrule
\end{tabular}
\end{table*}

Conveniently, of the five ML models introduced in Section~\ref{Sec:ML}, the out-of-the-box Scikit-learn implementation of SGD allows the linear model to be incrementally adjusted for new data via a `partial fit' method.
Accordingly, recall the best SGD models determined by HPO when trained on experimental or synthetic data pertaining to seven `already-encountered' SPS contexts; their hyperparameters are listed in Table~\ref{Tab:AdaptiveHPO}.
In the synthetic case, this table only lists models that are solely trained/validated on histograms of size $1000$ as -- this is shown statistically in Fig.~\ref{Fig:SynthAverage} -- such a data sample is of most benefit to early estimator SGD models.

Now, it is of course possible to simply treat both types of model, SGD-Exp and SGD-Syn, as static.
In such a case, they are iteratively fed histograms from an eighth unseen context and, at each step, make predictions of the best $g$ value.
The forecasts of SGD-Exp applied to the FI-SEQUR datasets are already shown in Fig.~\ref{Fig:ExperimentLinear}.
However, one can also incrementally predict and \textit{fit} on each newly encountered histogram, thus making the model adaptive.
The main problem is that the best value of $g$ is unknown for the eighth context, so the next best value that can be supplied at every ten-second timestep is the estimate made by the least-squares fitting procedure.
Note that, on most modern CPUs, using a single histogram to both perform least-squares fitting and apply `partial fit' to an SGD model takes far less than ten seconds, so this is easily an online process.

\begin{figure*}[p!]
\centering
\includegraphics[width=0.95\textwidth]{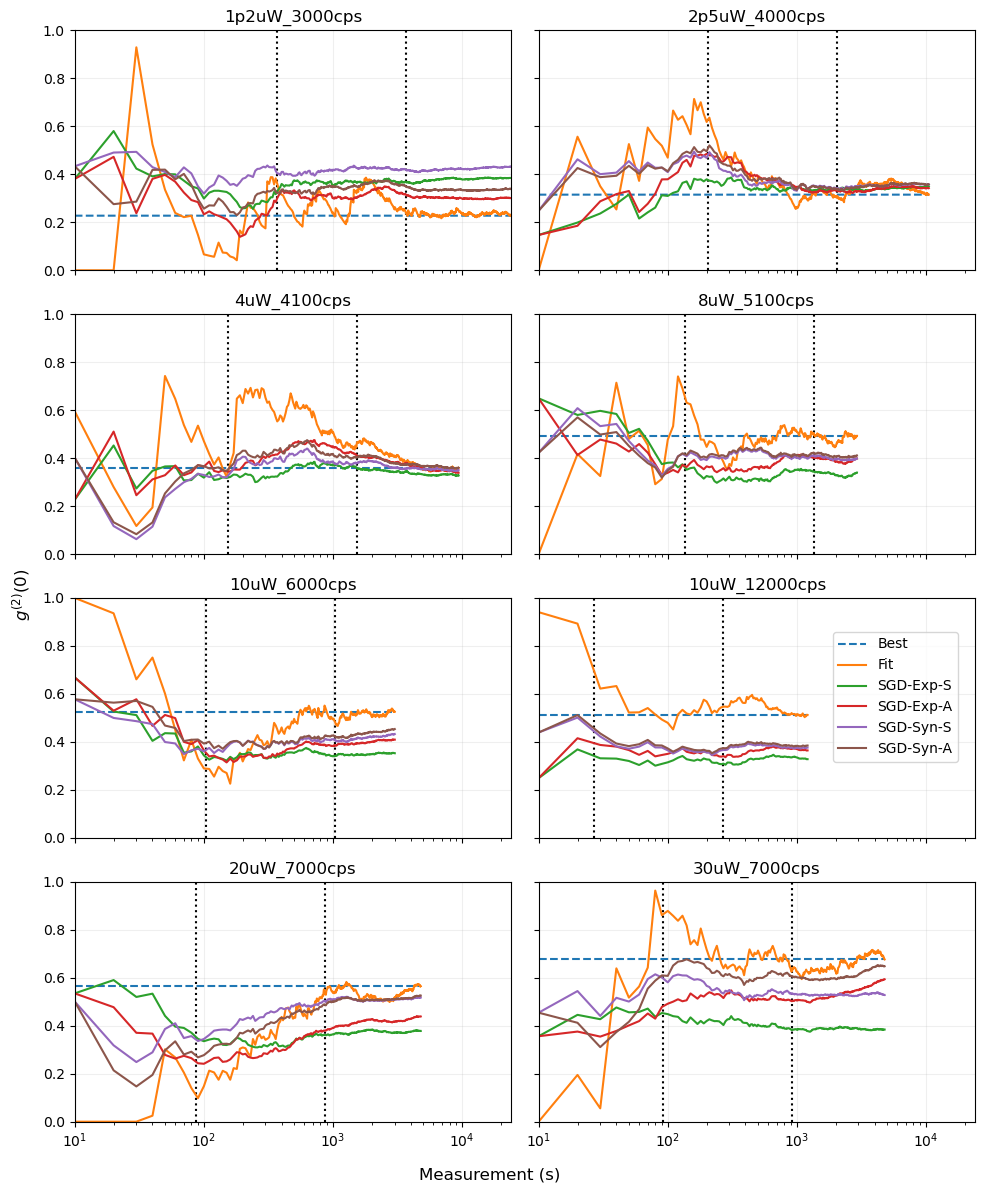}
\caption{\label{Fig:Adaptive} Fitting and static/adaptive SGD model estimates of $g$ for each FI-SEQUR dataset based on histograms of experimentally accumulated detections at every $10$ s timestep. The `Fit' trace denotes estimates made by least-squares fitting, with the final value defining the ground truth for each dataset, i.e.~a horizontal dashed `Best' line. Training is based on the other seven contexts: `SGD-Exp' for all available experimental histograms, and `SGD-Syn' for $14000$ synthetic size-$1000$ histograms. A static `S' model simply predicts, while an adaptive `A' model incrementally predicts and fits, using a learning rate of $0.2$ times the original training value in Table~\ref{Tab:AdaptiveHPO}. The measurement-time axis is logarithmic, and the first and second vertical dotted lines mark the times at which $1000$ and $10000$ detections are accumulated, respectively, per dataset.}
\end{figure*}

If this procedure for adaptation is followed, then it is possible to compare the forecasts of both static and adaptive models, suffixed here with S and A, respectively.
Indeed, the predictions made for the experimental FI-SEQUR datasets are plotted in Fig.~\ref{Fig:Adaptive}.
There is, however, a nuance to consider, i.e.~the strength of adaptation can be varied.
In the case of the Scikit-learn implementation, the default inverse scaling method means that earlier histograms are given stronger weighting than later histograms -- inverse scaling is typically useful for quick and substantial model adjustments followed by softening course corrections -- but the learning rate may need to be fine-tuned anyway.
Picking the best adaptive methodology is beyond the scope of this work; the adaptive models in Fig.~\ref{Fig:Adaptive} merely exemplify what happens when their original learning rates in Table~\ref{Tab:AdaptiveHPO} are multiplied by $0.2$ prior to encountering the new context.

\begin{table*}[htb!]
\centering
\caption{The loss values, i.e.~RMSE, for least-squares fitting and static/adaptive transfer-learning SGD models when tested on each context. Training is based on the other seven contexts: `SGD-Exp' for all available experimental histograms, and `SGD-Syn' for $14000$ synthetic size-$1000$ histograms. Testing is solely done on experimental data. A static `S' model simply predicts, while an adaptive `A' model incrementally predicts and fits, using a learning rate of $0.2$ times the original training value in Table~\ref{Tab:AdaptiveHPO}. The prediction RMSE is calculated for experimental histograms containing under $1000$ coincidences for `Early' loss, $1000$ to $10000$ coincidences for `Mid' loss, over $10000$ coincidences for `Late' loss, and any number of coincidences for `Full' loss. Loss is colour-coded: yellow to red for low to high values. Also included are the validation losses for SGD models, i.e.~the RMSE when applied to the seven contexts they were trained on; see Section~\ref{Sec:ML}. Note that training and validation on these seven contexts use independent subsamples of data, i.e.~no model is validated on data it is trained on.}
\label{Tab:AdaptiveLoss}
\begin{tabular}{llccccc}
\toprule
Context & Model & Loss (Early) & Loss (Mid) & Loss (Late) & Loss (Full) & Loss (Valid.) \\
\midrule\\[-3.5mm]
\multirow{5}{*}{1p2uW} & Fit & \cellcolor[rgb]{1.00,0.73,0.33} 0.1742 & \cellcolor[rgb]{1.00,0.91,0.59} 0.0769 & \cellcolor[rgb]{1.00,0.99,0.77} 0.0107 & \cellcolor[rgb]{1.00,0.96,0.70} 0.0371 &  \\
 & SGD-Exp-S & \cellcolor[rgb]{1.00,0.87,0.50} 0.1127 & \cellcolor[rgb]{1.00,0.82,0.43} 0.1404 & \cellcolor[rgb]{1.00,0.78,0.39} 0.1538 & \cellcolor[rgb]{1.00,0.79,0.39} 0.1514 & \cellcolor[rgb]{1.00,0.86,0.49} 0.1153 \\
 & SGD-Exp-A & \cellcolor[rgb]{1.00,0.91,0.59} 0.0781 & \cellcolor[rgb]{1.00,0.89,0.54} 0.0953 & \cellcolor[rgb]{1.00,0.92,0.60} 0.0730 & \cellcolor[rgb]{1.00,0.91,0.59} 0.0765 & \cellcolor[rgb]{1.00,0.86,0.49} 0.1153 \\
 & SGD-Syn-S & \cellcolor[rgb]{1.00,0.72,0.32} 0.1793 & \cellcolor[rgb]{1.00,0.69,0.29} 0.1931 & \cellcolor[rgb]{1.00,0.67,0.29} 0.1975 & \cellcolor[rgb]{1.00,0.68,0.29} 0.1966 & \cellcolor[rgb]{1.00,0.89,0.54} 0.0943 \\
 & SGD-Syn-A & \cellcolor[rgb]{1.00,0.89,0.54} 0.0971 & \cellcolor[rgb]{1.00,0.84,0.46} 0.1275 & \cellcolor[rgb]{1.00,0.87,0.51} 0.1081 & \cellcolor[rgb]{1.00,0.87,0.50} 0.1108 & \cellcolor[rgb]{1.00,0.89,0.54} 0.0943 \\
\cline{1-7}\\[-3.5mm]
\multirow{5}{*}{2p5uW} & Fit & \cellcolor[rgb]{0.99,0.45,0.21} 0.2755 & \cellcolor[rgb]{1.00,0.92,0.62} 0.0669 & \cellcolor[rgb]{1.00,0.96,0.70} 0.0356 & \cellcolor[rgb]{1.00,0.94,0.64} 0.0570 &  \\
 & SGD-Exp-S & \cellcolor[rgb]{1.00,0.92,0.62} 0.0683 & \cellcolor[rgb]{1.00,0.97,0.72} 0.0283 & \cellcolor[rgb]{1.00,0.97,0.72} 0.0277 & \cellcolor[rgb]{1.00,0.97,0.72} 0.0291 & \cellcolor[rgb]{1.00,0.93,0.63} 0.0617 \\
 & SGD-Exp-A & \cellcolor[rgb]{1.00,0.87,0.49} 0.1150 & \cellcolor[rgb]{1.00,0.93,0.64} 0.0587 & \cellcolor[rgb]{1.00,0.97,0.72} 0.0276 & \cellcolor[rgb]{1.00,0.96,0.70} 0.0383 & \cellcolor[rgb]{1.00,0.93,0.63} 0.0617 \\
 & SGD-Syn-S & \cellcolor[rgb]{1.00,0.83,0.44} 0.1344 & \cellcolor[rgb]{1.00,0.94,0.66} 0.0494 & \cellcolor[rgb]{1.00,0.96,0.69} 0.0404 & \cellcolor[rgb]{1.00,0.95,0.68} 0.0457 & \cellcolor[rgb]{1.00,0.91,0.59} 0.0791 \\
 & SGD-Syn-A & \cellcolor[rgb]{1.00,0.81,0.42} 0.1412 & \cellcolor[rgb]{1.00,0.93,0.64} 0.0589 & \cellcolor[rgb]{1.00,0.96,0.69} 0.0410 & \cellcolor[rgb]{1.00,0.95,0.67} 0.0484 & \cellcolor[rgb]{1.00,0.91,0.59} 0.0791 \\
\cline{1-7}\\[-3.5mm]
\multirow{5}{*}{4uW} & Fit & \cellcolor[rgb]{1.00,0.72,0.32} 0.1778 & \cellcolor[rgb]{1.00,0.69,0.29} 0.1919 & \cellcolor[rgb]{1.00,0.96,0.69} 0.0394 & \cellcolor[rgb]{1.00,0.90,0.57} 0.0850 &  \\
 & SGD-Exp-S & \cellcolor[rgb]{1.00,0.94,0.64} 0.0569 & \cellcolor[rgb]{1.00,0.98,0.75} 0.0190 & \cellcolor[rgb]{1.00,0.97,0.74} 0.0237 & \cellcolor[rgb]{1.00,0.97,0.74} 0.0240 & \cellcolor[rgb]{1.00,0.92,0.60} 0.0738 \\
 & SGD-Exp-A & \cellcolor[rgb]{1.00,0.93,0.63} 0.0628 & \cellcolor[rgb]{1.00,0.91,0.58} 0.0809 & \cellcolor[rgb]{1.00,0.98,0.75} 0.0204 & \cellcolor[rgb]{1.00,0.96,0.70} 0.0371 & \cellcolor[rgb]{1.00,0.92,0.60} 0.0738 \\
 & SGD-Syn-S & \cellcolor[rgb]{1.00,0.84,0.46} 0.1270 & \cellcolor[rgb]{1.00,0.96,0.70} 0.0380 & \cellcolor[rgb]{1.00,0.99,0.77} 0.0106 & \cellcolor[rgb]{1.00,0.97,0.74} 0.0237 & \cellcolor[rgb]{1.00,0.91,0.59} 0.0772 \\
 & SGD-Syn-A & \cellcolor[rgb]{1.00,0.87,0.49} 0.1138 & \cellcolor[rgb]{1.00,0.92,0.62} 0.0674 & \cellcolor[rgb]{1.00,0.98,0.75} 0.0201 & \cellcolor[rgb]{1.00,0.96,0.71} 0.0348 & \cellcolor[rgb]{1.00,0.91,0.59} 0.0772 \\
\cline{1-7}\\[-3.5mm]
\multirow{5}{*}{8uW} & Fit & \cellcolor[rgb]{1.00,0.68,0.29} 0.1967 & \cellcolor[rgb]{1.00,0.95,0.68} 0.0434 & \cellcolor[rgb]{1.00,0.98,0.75} 0.0194 & \cellcolor[rgb]{1.00,0.94,0.66} 0.0521 &  \\
 & SGD-Exp-S & \cellcolor[rgb]{1.00,0.88,0.53} 0.0993 & \cellcolor[rgb]{1.00,0.76,0.36} 0.1622 & \cellcolor[rgb]{1.00,0.76,0.37} 0.1615 & \cellcolor[rgb]{1.00,0.77,0.37} 0.1595 & \cellcolor[rgb]{1.00,0.91,0.59} 0.0792 \\
 & SGD-Exp-A & \cellcolor[rgb]{1.00,0.87,0.50} 0.1127 & \cellcolor[rgb]{1.00,0.87,0.50} 0.1115 & \cellcolor[rgb]{1.00,0.88,0.52} 0.1034 & \cellcolor[rgb]{1.00,0.87,0.51} 0.1073 & \cellcolor[rgb]{1.00,0.91,0.59} 0.0792 \\
 & SGD-Syn-S & \cellcolor[rgb]{1.00,0.88,0.53} 0.0995 & \cellcolor[rgb]{1.00,0.90,0.58} 0.0835 & \cellcolor[rgb]{1.00,0.89,0.54} 0.0958 & \cellcolor[rgb]{1.00,0.89,0.55} 0.0910 & \cellcolor[rgb]{1.00,0.91,0.59} 0.0773 \\
 & SGD-Syn-A & \cellcolor[rgb]{1.00,0.88,0.53} 0.1004 & \cellcolor[rgb]{1.00,0.91,0.60} 0.0749 & \cellcolor[rgb]{1.00,0.90,0.57} 0.0842 & \cellcolor[rgb]{1.00,0.91,0.58} 0.0812 & \cellcolor[rgb]{1.00,0.91,0.59} 0.0773 \\
\cline{1-7}\\[-3.5mm]
\multirow{5}{*}{10uW--} & Fit & \cellcolor[rgb]{0.99,0.56,0.24} 0.2495 & \cellcolor[rgb]{1.00,0.88,0.53} 0.0978 & \cellcolor[rgb]{1.00,0.98,0.76} 0.0172 & \cellcolor[rgb]{1.00,0.92,0.60} 0.0725 &  \\
 & SGD-Exp-S & \cellcolor[rgb]{1.00,0.84,0.46} 0.1274 & \cellcolor[rgb]{1.00,0.73,0.33} 0.1765 & \cellcolor[rgb]{1.00,0.73,0.33} 0.1762 & \cellcolor[rgb]{1.00,0.73,0.33} 0.1749 & \cellcolor[rgb]{1.00,0.91,0.60} 0.0748 \\
 & SGD-Exp-A & \cellcolor[rgb]{1.00,0.88,0.53} 0.1007 & \cellcolor[rgb]{1.00,0.77,0.38} 0.1569 & \cellcolor[rgb]{1.00,0.85,0.46} 0.1257 & \cellcolor[rgb]{1.00,0.83,0.44} 0.1356 & \cellcolor[rgb]{1.00,0.91,0.60} 0.0748 \\
 & SGD-Syn-S & \cellcolor[rgb]{1.00,0.86,0.47} 0.1221 & \cellcolor[rgb]{1.00,0.85,0.46} 0.1267 & \cellcolor[rgb]{1.00,0.87,0.51} 0.1072 & \cellcolor[rgb]{1.00,0.87,0.49} 0.1141 & \cellcolor[rgb]{1.00,0.91,0.59} 0.0794 \\
 & SGD-Syn-A & \cellcolor[rgb]{1.00,0.90,0.56} 0.0861 & \cellcolor[rgb]{1.00,0.86,0.48} 0.1188 & \cellcolor[rgb]{1.00,0.90,0.56} 0.0885 & \cellcolor[rgb]{1.00,0.88,0.53} 0.0989 & \cellcolor[rgb]{1.00,0.91,0.59} 0.0794 \\
\cline{1-7}\\[-3.5mm]
\multirow{5}{*}{10uW+} & Fit & \cellcolor[rgb]{0.82,0.05,0.13} 0.4052 & \cellcolor[rgb]{1.00,0.95,0.68} 0.0434 & \cellcolor[rgb]{1.00,0.96,0.70} 0.0366 & \cellcolor[rgb]{1.00,0.93,0.63} 0.0643 &  \\
 & SGD-Exp-S & \cellcolor[rgb]{0.99,0.64,0.27} 0.2121 & \cellcolor[rgb]{1.00,0.69,0.29} 0.1920 & \cellcolor[rgb]{1.00,0.71,0.31} 0.1817 & \cellcolor[rgb]{1.00,0.71,0.31} 0.1843 & \cellcolor[rgb]{1.00,0.94,0.66} 0.0511 \\
 & SGD-Exp-A & \cellcolor[rgb]{1.00,0.67,0.29} 0.1985 & \cellcolor[rgb]{1.00,0.77,0.38} 0.1571 & \cellcolor[rgb]{1.00,0.80,0.40} 0.1467 & \cellcolor[rgb]{1.00,0.79,0.40} 0.1498 & \cellcolor[rgb]{1.00,0.94,0.66} 0.0511 \\
 & SGD-Syn-S & \cellcolor[rgb]{1.00,0.94,0.66} 0.0523 & \cellcolor[rgb]{1.00,0.80,0.41} 0.1460 & \cellcolor[rgb]{1.00,0.83,0.44} 0.1331 & \cellcolor[rgb]{1.00,0.83,0.44} 0.1349 & \cellcolor[rgb]{1.00,0.91,0.58} 0.0803 \\
 & SGD-Syn-A & \cellcolor[rgb]{1.00,0.94,0.66} 0.0517 & \cellcolor[rgb]{1.00,0.82,0.43} 0.1379 & \cellcolor[rgb]{1.00,0.85,0.47} 0.1249 & \cellcolor[rgb]{1.00,0.85,0.46} 0.1268 & \cellcolor[rgb]{1.00,0.91,0.58} 0.0803 \\
\cline{1-7}\\[-3.5mm]
\multirow{5}{*}{20uW} & Fit & \cellcolor[rgb]{0.64,0.00,0.15} 0.4632 & \cellcolor[rgb]{1.00,0.69,0.29} 0.1931 & \cellcolor[rgb]{1.00,0.96,0.70} 0.0364 & \cellcolor[rgb]{1.00,0.88,0.52} 0.1038 &  \\
 & SGD-Exp-S & \cellcolor[rgb]{1.00,0.86,0.48} 0.1207 & \cellcolor[rgb]{0.99,0.61,0.26} 0.2268 & \cellcolor[rgb]{1.00,0.69,0.30} 0.1910 & \cellcolor[rgb]{1.00,0.68,0.29} 0.1963 & \cellcolor[rgb]{1.00,0.89,0.55} 0.0931 \\
 & SGD-Exp-A & \cellcolor[rgb]{0.99,0.59,0.25} 0.2328 & \cellcolor[rgb]{0.99,0.56,0.24} 0.2471 & \cellcolor[rgb]{1.00,0.80,0.41} 0.1454 & \cellcolor[rgb]{1.00,0.74,0.35} 0.1680 & \cellcolor[rgb]{1.00,0.89,0.55} 0.0931 \\
 & SGD-Syn-S & \cellcolor[rgb]{0.99,0.62,0.27} 0.2200 & \cellcolor[rgb]{1.00,0.87,0.50} 0.1119 & \cellcolor[rgb]{1.00,0.94,0.65} 0.0549 & \cellcolor[rgb]{1.00,0.92,0.60} 0.0730 & \cellcolor[rgb]{1.00,0.91,0.59} 0.0786 \\
 & SGD-Syn-A & \cellcolor[rgb]{0.99,0.36,0.18} 0.2999 & \cellcolor[rgb]{1.00,0.79,0.39} 0.1511 & \cellcolor[rgb]{1.00,0.94,0.65} 0.0534 & \cellcolor[rgb]{1.00,0.90,0.56} 0.0870 & \cellcolor[rgb]{1.00,0.91,0.59} 0.0786 \\
\cline{1-7}\\[-3.5mm]
\multirow{5}{*}{30uW} & Fit & \cellcolor[rgb]{0.90,0.12,0.11} 0.3697 & \cellcolor[rgb]{1.00,0.94,0.65} 0.0553 & \cellcolor[rgb]{1.00,0.96,0.71} 0.0321 & \cellcolor[rgb]{1.00,0.93,0.63} 0.0627 &  \\
 & SGD-Exp-S & \cellcolor[rgb]{0.99,0.59,0.25} 0.2363 & \cellcolor[rgb]{0.99,0.47,0.21} 0.2700 & \cellcolor[rgb]{0.99,0.39,0.19} 0.2898 & \cellcolor[rgb]{0.99,0.41,0.19} 0.2856 & \cellcolor[rgb]{1.00,0.89,0.54} 0.0949 \\
 & SGD-Exp-A & \cellcolor[rgb]{0.99,0.46,0.21} 0.2742 & \cellcolor[rgb]{1.00,0.76,0.37} 0.1608 & \cellcolor[rgb]{1.00,0.84,0.45} 0.1321 & \cellcolor[rgb]{1.00,0.81,0.42} 0.1414 & \cellcolor[rgb]{1.00,0.89,0.54} 0.0949 \\
 & SGD-Syn-S & \cellcolor[rgb]{1.00,0.78,0.38} 0.1548 & \cellcolor[rgb]{1.00,0.83,0.44} 0.1338 & \cellcolor[rgb]{1.00,0.80,0.41} 0.1458 & \cellcolor[rgb]{1.00,0.81,0.41} 0.1440 & \cellcolor[rgb]{1.00,0.90,0.56} 0.0886 \\
 & SGD-Syn-A & \cellcolor[rgb]{0.99,0.60,0.25} 0.2312 & \cellcolor[rgb]{1.00,0.93,0.62} 0.0648 & \cellcolor[rgb]{1.00,0.93,0.63} 0.0624 & \cellcolor[rgb]{1.00,0.92,0.61} 0.0699 & \cellcolor[rgb]{1.00,0.90,0.56} 0.0886 \\
\cline{1-7}
\bottomrule
\end{tabular}
\end{table*}

Naturally, as discussed before, even the FI-SEQUR datasets provide limited experimental data to draw large-scale statistical conclusions.
However, the results in Fig.~\ref{Fig:Adaptive} and the Table~\ref{Tab:AdaptiveLoss} breakdown suggest, first, that training on size-$1000$ histograms has the intended like-for-like benefit.
Examination of the traces at the first dotted vertical line, at which point $1000$ coincidences have been detected, shows SGD-Syn-S closer than SGD-Exp-S to the goal in four contexts and on par in another two.
The models trained on synthetic data also seem to do better in the long term for higher laser intensities, possibly because their training data is equally weighted across seven contexts.
Specifically, most SGD-Exp models are overly influenced by the 1p2uW context as that FI-SEQUR dataset is the largest.

The results also show that adaptation indeed has its effect, with the SGD-Exp-A and SGD-Syn-A traces being pulled towards the least-squares fitting estimates.
That stated, it is an open question as to whether chasing such estimates, especially in early stages when they are subject to the most stochastic error, is a beneficial exercise.
The medium-term predictions in both the 20uW and 30uW contexts, i.e.~loss (mid), reveal where this form of adaptation is negative and positive, respectively, and they depend solely on whether least-squares fitting is presently doing worse or better than transfer ML.
Granted, this simplistic behaviour is due to a basic form of adaptation applied to a linear model.
In theory, ensemble models can more effectively weigh the influence of new data, e.g.~via the consensus of multiple learners, but this is beyond the scope of the current work.

\section{Discussion}
\label{Sec:Discussion}

In many applications, ML provides a benefit through fast inference.
While HPO can take hours or even days to train and validate numerous candidate models, the resulting object usually converts an input feature space into an output prediction within a matter of seconds.
However, in the case of early estimation for SPS quality, least-squares fitting of Eq.~(\ref{Eq:Fit}) is already formidably efficient, especially given the alternative form of the physically motivated function~\cite{kemu23}.
Indeed, both fitting and model inference can easily be done in streamed fashion for an HBT interferometry experiment with ten-second increments.
Thus, the utility of ML will ultimately come down to its relative accuracy.

Validation results in Section~\ref{Sec:Experiment} already indicate that the analytical OLS and the more complex ensemble models can easily pick up patterns within coincidence histograms that least-squares fitting cannot identify and leverage; SGD and PLSR do not do poorly either, although the former is dependent on good HPO and the latter relies on clear covariances existing within training data.
So, the learning capability of ML is not in question.
The problem is that the prospective use case of ML in SPS quality estimation requires the models to be transferred from one context to another.
Unfortunately, discounting the relative statistics when averaged across all FI-SEQUR contexts, i.e.~Fig.~\ref{Fig:SynthAverage}, the long-term behaviour and the prediction averages of the transferred models, e.g.~those shown in Fig.~\ref{Fig:ExperimentLinear} and Fig.~\ref{Fig:SynthOLS}, are somewhat unconvincing as to their general applicability.

So, does a variation in laser intensity really make a fundamental physical difference to the cumulative histograms of detections?
Quite possibly.
When laser intensities are low, the background may be dominated by thermal noise within the detectors, i.e.~so-called dark counts.
When laser intensities are high, the excess energy can contribute to other significant sources of sample-related background.
Essentially, ML model inaccuracies at the extremes of laser intensity may not just be due to an averaging effect; the models may simply not be able to extrapolate for the physics of these disparate contexts.
In fact, in the field of ML, the subtopic of meta-learning \cite{lebu15,albu20} often grapples with the conundrum of context similarity, trying to quantify `distances' between ML problems that inversely correspond with the transferability of ML techniques.
Such a metric in the SPS case may well depend on the other four characteristic values in Table~\ref{Tab:BestFit} beyond $g$, and this is already assuming the nature of both the quantum dot and the stimulating laser are fixed.
Quantifying context similarity is a hard problem.

That stated, if the transferability of the studied ML models is suffering due to hidden variables, possibly nonlinear, there are ways forward.
First of all, the five models studied in this investigation were split between linearly simple and nonlinearly complex.
Specifically, RF and GB were chosen because the two are strong performers in a variety of ML problems.
That does not mean they are automatically suitable for the SPS case, and they were indeed very slow to tune via HPO on local hardware.
Given the diversity of ML models in existence, a simpler but still nonlinear predictor may be able to convincingly extrapolate into new contexts where OLS struggles.

Then there are the features to consider, namely the $1954$ coincidence counts across the range of the histogram.
Many of these are likely to be redundant, at least when considered independently rather than immediately averaged.
After all, the value of $g$ is effectively the amplitude of one peak divided by the amplitude of any other, both adjusted by the background.
So, while the way a histogram evolves in its early stages may statistically indicate the eventual values for all three metrics of importance, $1954$ independent features may confound ML more than it informs.
Feature selection, e.g.~principle component analysis, is one way to shrink this set, hopefully without sacrificing information content.
Of course, if linear operations applied to the $1954$ detection counts cannot capture nonlinear context variations, the other avenue is also to include more features, e.g.~spectral information, or generate new ones from the ones that currently exist, e.g.~via simple transforms like Fourier decomposition.
This is called feature engineering.

Finally, when all else fails to prepare for a new context, adaptation is an avenue to adjust an ML model.
For instance, consider a new SPS context with wider peaks and a smaller $\gamma_p$ decay factor, for which early-stage HBT interferometry would probably witness a loose clustering of detected coincidences near those peaks.
Least-squares fitting would likely interpret those detections as shaping part of the peaks, while a transferred ML model might incorrectly dismiss the phenomenon as part of background rate $R_b$.
Thus, adaptation could update the internal parameters of the model to account for this contextual shift.
However, while this work has shown adaptation to work in principle, it remains an open question as to its practical use.
Adaptation is often a gradual process, so as not to catastrophically interfere with pre-existing knowledge, while effective early estimation requires rapid adjustment.
Future work will also need to decide how best to leverage evolving least-squares estimates, short of finding a better ground-truth proxy, without merely `chasing' these predictions as if they were better.

As a concluding comment, while the topic of SPS evaluation has, up to now, remained almost untouched by ML research, the impetus to cross-polinate methodologies is growing.
For instance, a very recent publication has sought to use an autoencoding convolutional neural network to predict the SPS quality of candidate quantum dots~\cite{coja24}.
It differs from this work in that it primarily uses emission spectra rather than second-order correlation histograms as the inputs for its ML model, and it relies on expert labelling rather than an automatically calculated $g$ metric for its target values.
In effect, the use of ML in the publication appears to avoid costly HBT interferometry outright, albeit by instead incorporating subjectivity within its model.
There are pros and cons to either approach, but this goes to show that the promise of ML methodologies for SPS quality estimation is not ignorable, and further research in this space should be expected.

\section{Conclusion}
\label{Sec:Conclusion}

The degree to which a quantum dot can be considered an SPS depends on how frequently any two emitted photons are determined to have been released synchronously.
Unfortunately, detecting enough coincidences to build up a second-order correlation function for these emission statistics, especially for certain levels of confidence, takes significant time and material resources.
For this reason, there are ongoing research efforts to minimise such costs by exploring early estimation of SPS quality, which, with some modifications, can also be extended to other correlation experiments.

At the same time, a growing understanding and appreciation of ML with its potential for inductive learning and prediction has seen associated techniques increasingly applied to the physical sciences.
Given that the particular niche of early estimation for SPS quality is both fertile ground and essentially untouched by ML research, this work aimed to amend that by running a novel investigation into whether ML techniques could do better than traditional least-squares fitting of a physically motivated function.
The research was supported by a large collection of clean and easily comparable experimental data, namely eight FI-SEQUR datasets, which detail the accumulation of detected photon coincidences for a fibre-coupled InGaAs/GaAs epitaxial quantum dot subject to varying pulsed-laser intensities.

In detail, for each of these eight SPS `contexts', ML models were first trained on either purely experimental or synthetically generated data from the other seven contexts; the data consisted of normalised coincidence histograms as inputs and best least-squares estimates for $g^{(2)}(0)$ -- here called `ground truths' -- as target values.
The models were then tested on the eighth context against estimates provided by standard least-squares fitting.
Three model types were linear, i.e.~OLS/SGD/PLSR, and two were ensemble-based, i.e.~RF/GB.
Their hyperparameters, including one determining the use of a standard scaling preprocessor, were selected by bandit-based Bayesian optimisation, where applicable, prior to full training.

It was found that:
\begin{itemize}
\item Standard ML can indeed pick up patterns between histograms and ground truths that standard least-squares fitting ignores.
Indeed, validation scores indicated the superiority of ML for OLS and the ensemble models, with SGD and PLSR not performing badly either.
Thus, if an ML model were to be applied to the same contexts it was trained on, i.e.~the theoretical extreme of context similarity, it would not be difficult to improve over current practices for quality estimation.

\item Transfer learning, where an ML model is tested in a new previously unseen context, is a much more tricky proposition and, unfortunately, is presently the more realistic use case for ML.
Nonetheless, thanks to realistically sampled synthetically generated data, statistical inferences could be made that would be unavailable with purely experimental data.
In the case of the FI-SEQUR contexts, all five ML models, on average, performed better than least-squares fitting as an early estimator, i.e.~at ${\sim}1000$ coincidences detected.
For OLS and SGD, there was a further `like-for-like' implication, i.e.~models should train on histograms similarly sized to those they are applied to.
However, further research is required to ascertain how generalisable this result is.

\item Adaptation can be used to tweak an ML model in response to new data from an unseen context, and this was showcased by SGD.
However, without a known ground truth, one must choose a proxy, e.g.~the best least-squares estimate of SPS quality thus far.
Combined with the simple form of adaptation built into the linear model, this manifested as SGD predictions `chasing' least-squares predictions, which is only useful if the ML model is doing worse in the first place.
More sophisticated adaptation methods could allow ML models to be influenced more subtly, but this too needs additional investigation.
\end{itemize}

In summary, this work has shown ML has strong potential to improve upon standard fitting methods for SPS quality estimation, but the transferability between contexts remains a conundrum.
Varying laser intensity was enough to noticeably damage performance for the ML models studied, even if the alternative proved statistically weaker, and this stymies generalising transfer learning across greater context variations, e.g.~different quantum dots and laser profiles.
Future paths for progress may involve experimenting with feature selection/generation, i.e.~to refine the coincidence histograms and find informative nonlinearities, as well as isolating conflating factors within SPS contexts, e.g.~varying rates of background detections.
Of course, there are also many other predictive ML models to leverage as well, each with varying complexity and applicability, and some with adaptive measures that can pivot to information from new contexts.
Ideally, the conclusions of this investigation highlight the challenges of transfer learning and where best to focus efforts, so that the power of ML can eventually be deployed effectively for the early estimation of SPS quality.

\begin{acknowledgments}
The experimental data used in this study was obtained within the FI-SEQUR project, which was jointly financed by (1) the European Regional Development Fund (EFRE) of the European Union, as part of a programme to promote research, innovation, and technology (Pro FIT) in Germany, and (2) the National Centre for Research and Development in Poland, as part of the 2nd Poland-Berlin Photonics Programme, grant number 2/POLBER-2/2016 (project value: 2089498 PLN). 
\end{acknowledgments}

\section*{Author Declarations}
\subsection*{Conflict of Interest}
The authors have no conflicts to disclose.

\subsection*{Author Contributions}
\textbf{David Jacob Kedziora}: Conceptualisation (supporting); Formal Analysis (lead); Methodology (lead); Software (lead); Visualisation (lead); Writing -- original draft (lead); Writing -- review \& editing (equal). \textbf{Anna Musiał}: Conceptualisation (equal); Data Curation (lead); Investigation (equal); Writing -- review \& editing (equal). \textbf{Wojciech Rudno-Rudziński}: Investigation (equal); Writing -- review \& editing (equal). \textbf{Bogdan Gabrys}: Conceptualisation (equal); Methodology (supporting); Project Administration (lead); Supervision (lead); Writing -- review \& editing (equal).

\section*{Data Availability}
The data that supports the findings of this study is openly available at \href{https://github.com/UTS-CASLab/sps-quality}{github.com/UTS-CASLab/sps-quality}.

\section*{References}
\bibliography{main}% Produces the bibliography via BibTeX.

\end{document}